\documentclass{aa}
\usepackage{graphicx}
\usepackage{amssymb}
\usepackage{amsmath}
\usepackage[varg]{txfonts}

\usepackage{natbib}
\usepackage{xspace}
\usepackage{xcolor}
\usepackage{hyperref}

\hypersetup{
    hidelinks,
    colorlinks = false
}

\bibpunct{(}{)}{;}{a}{}{,}

\makeatletter
\renewcommand*\aa@pageof{, page \thepage{} of \pageref*{LastPage}}
\makeatother

\newcommand\incl{\ensuremath{i}\xspace}
\newcommand\Kalpha{K\ensuremath{\alpha}\xspace}
\newcommand\rg{\ensuremath{r_\mathrm{g}}\xspace}
\newcommand\reh{\ensuremath{r_\mathrm{EH}}\xspace}
\newcommand\risco{\ensuremath{r_\mathrm{ISCO}}\xspace}
\newcommand\rin{\ensuremath{r_\mathrm{in}}\xspace}
\newcommand\rout{\ensuremath{r_\mathrm{out}}\xspace}

\newcommand\ringradius{\ensuremath{x}\xspace}
\newcommand\diskradius{\ensuremath{r_\mathrm{d}}\xspace}
\newcommand\sphericalradius{\ensuremath{r}\xspace}
\newcommand\Kerrarea{\ensuremath{A_\mathrm{Kerr}}\xspace}
\newcommand\flatarea{\ensuremath{A_\mathrm{flat}}\xspace}
\newcommand\fluxspecific{\ensuremath{\Phi}\xspace}

\newcommand\xmm{\textsl{XMM-Newton}\xspace}
\newcommand\nustar{\textsl{NuSTAR}\xspace}

\newcommand\ynogk{\textsc{ynogk}\xspace}

\newcommand\xillvercp{\textsc{xillver\_cp}\xspace}
\newcommand\relxill{\textsc{relxill}\xspace}

\newcommand\relxilllpext{\textsc{relxill\_ring}\xspace} % was relxill_ext in v2.6, in v2.7 rename to relxill_ring
\newcommand\relxilllpcp{\textsc{relxilllp\_cp}\xspace}
\newcommand\apec{\textsc{apec}\xspace}
\newcommand\nthcomp{\textsc{nthcomp}\xspace}
\newcommand\nthcompscat{\textsc{nthcomp}$_\mathrm{scat}$\xspace}
\newcommand\tbabs{\textsc{tbabs}\xspace}
\newcommand\tbnewfeo{\textsc{tbnew\_feo}\xspace}
\newcommand\xstar{\textsc{xstar}\xspace}
\newcommand\detconst{\textsc{detconst}\xspace}

\newcommand\Waltontwentyone{W21\xspace}
\newcommand\esosource{ESO\,033-G002\xspace}
\newcommand\modellp{Model~1b\xspace}
\newcommand\modellpold{Model~1a\xspace}
\newcommand\modelextfixboost{Model~2b\xspace}
\newcommand\modelextfreeboost{Model~2a\xspace}

\newcommand\norm{\ensuremath{\mathrm{Norm}}\xspace}
\newcommand\NH{\ensuremath{N_{\mathrm{H}}}\xspace}
\newcommand\covfrac{\ensuremath{C_{\mathrm{f}}}\xspace}
\newcommand\logxi{\ensuremath{\log\xi}\xspace}
\newcommand\logdensity{\ensuremath{\log n}\xspace}
\newcommand\vout{\ensuremath{v_{\mathrm{out}}}\xspace}
\newcommand\kT{\ensuremath{k T}\xspace}
\newcommand\kTe{\ensuremath{k T_{\mathrm{e}}}\xspace}
\newcommand\reflfrac{\ensuremath{R_{\mathrm{frac}}}\xspace}
\newcommand\feabund{\ensuremath{A_{\mathrm{Fe}}}\xspace}
\newcommand\boost{\ensuremath{\textsc{boost}}\xspace}

\newcommand\cmminustwo{\ensuremath{\mathrm{cm}^{-2}}}
\newcommand\cmminusthree{\ensuremath{\mathrm{cm}^{-3}}}

\newcommand{\dif}{\ensuremath{\mathrm{d}}\xspace}

\newcommand{\OmegaSource}{\ensuremath{\Omega}\xspace}
\newcommand{\OmegaKeplerian}{\ensuremath{\Omega_\mathrm{K}}\xspace}

\newcommand{\Edisk}{\ensuremath{E_\mathrm{disk}}\xspace}
\newcommand{\Esource}{\ensuremath{E_\mathrm{source}}\xspace}

\newcommand{\Erest}{\ensuremath{E_\mathrm{rest}}\xspace}
\newcommand{\Elnrf}{\ensuremath{E_\mathrm{LNRF}}\xspace}
\newcommand{\Einf}{\ensuremath{E_\mathrm{inf}}\xspace}

\newcommand{\fractiondisk}{\ensuremath{q_\mathrm{disk}}\xspace}
\newcommand{\fractionesc}{\ensuremath{q_\mathrm{esc}}\xspace}
\newcommand{\fractionblackhole}{\ensuremath{q_\mathrm{bh}}\xspace}

\graphicspath{{./figures/}}

\begin{document}

   \title{Relativistic reflection within an extended \\ hot plasma geometry}

   \author{{A.\ D.\ Nekrasov\inst{1}}
          \and
          T.\ Dauser\inst{1}
          \and
          J.\ A.\ Garc\'ia\inst{2,3}
          \and
          D.\ J.\ Walton\inst{4}
          \and
          C.\ M.\ Fromm\inst{5,6,7}
          \and
          A.\ J.\ Young\inst{8}
          \and
          F.\ J.\ E.\ Baker\inst{8}
          \and
          A.\ M.\ Joyce\inst{1}
          \and
          {O.\ K\"onig\inst{9}}
          \and
          S.\ Licklederer\inst{1}
          \and
          J.\ H\"afner\inst{1}
          \and
          J.\ Wilms\inst{1}
          }

   \institute{Dr.\ Karl Remeis-Sternwarte~\&~ECAP, Friedrich-Alexander-Universit\"at Erlangen-N\"urnberg, Sternwartstr.~7, 96049 Bamberg, Germany \and
    NASA Goddard Space Flight Center, X-ray Astrophysics Laboratory, 8800 Greenbelt Road, Greenbelt, MD 20771, USA \and
    Cahill Center for Astrophysics, California Institute of Technology, 1216 East California Boulevard, Pasadena, CA 91125, USA \and
    Centre for Astrophysics Research, University of Hertfordshire, College Lane, Hatfield AL10 9AB, United Kingdom \and
    Institut f\"ur Theoretische Physik und Astrophysik, Universit\"at W\"urzburg, Emil-Fischer-Str.~31, 97074 W\"urzburg, Germany \and
    Institut f\"ur Theoretische Physik, Goethe Universit\"at, Max-von-Laue-Str.~1, 60438 Frankfurt, Germany \and
    Max-Planck-Institut f\"ur Radioastronomie, Auf dem H\"ugel 69, 53121 Bonn, Germany \and
    H.\ H.\ Wills Physics Laboratory, Tyndall Avenue, Bristol BS8 1TL, United Kingdom
    \and
    Center for Astrophysics | Harvard \& Smithsonian, 60 Garden Street, Cambridge, MA 02138, USA \\
    \email{alex.nekrasov@fau.de}
    }

   \date{Received June 18, 2025; accepted September 27, 2025}

  \abstract
    {The reflection of X-rays at the inner accretion disk around black holes imprints relativistically broadened features in the observed spectrum. Aside from the black hole properties and the ionization and density of the accretion disk, these features also depend on the location and geometry of the primary source of X-rays, often referred to as the corona.}
   {We present a fast general relativistic model for spectral fitting of a radially extended, ring-like corona above the accretion disk.}
   {A common approach used to explain observed X-ray reflection spectra is the lamp post geometry, which assumes a point-like source on the rotational axis of the black hole. While it is typically able to explain the observations, this geometric model does not allow for any constraint to be placed on the radial size of the corona. We therefore extended the publicly available relativistic reflection model \relxill by implementing a radially extended, ring-like primary source.}
   {With the new \relxill model allowing us to vary the position of the primary source in two dimensions, we present simulated line profiles and spectra and discuss the implications of carrying out a data fitting, in comparison to the lamp post model. We applied this extended \relxill model to \xmm and \nustar data of the radio-quiet Seyfert-2 active galactic nucleus (AGN) ESO\,033-G002. The new model describes the data well and we are able to constrain the distance of the source to the black hole to be less than three gravitational radii, while the angular position of the source is poorly constrained.}
   {We show that a compact, radially extended corona close to the innermost stable circular orbit is able to explain the observed relativistic reflection as well as the % point-like
   lamp post corona does. This model has been made freely available to the community.}


   \maketitle   

\section{Introduction}   \label{sec:Intro}
X-rays carry direct information from the innermost region around black holes (BHs). Thermal radiation from an accretion disk \citep{shakura1973a} is upscattered by % an % (wrong) 
inverse Compton effect in a hot plasma, often called the ``corona,'' which is located close to the BH \citep{sunyaev1980,haardt1993}. Some of these hard X-rays then irradiate the disk, where they are reprocessed in a process called reflection \citep{george1991, matt1993, krolik1994, blackman1999, ross2005, garcia2010}. The strongest feature of the reflection spectrum is typically the fluorescent Fe~\Kalpha line at 6.4\,keV, while the detailed shape of spectrum depends on the parameters of the accretion disk such as its ionization and density \citep{garcia2016}. 

If reflection takes place in the vicinity of the BH, strong general relativistic energy shifts and special relativistic Doppler boosting affect the spectral shape \citep{lightman1988,fabian1989,tanaka1995,reynolds2003,dauser2010}. These effects depend on the observer's viewing angle and carry information about the matter and space-time in the vicinity of the BH \citep[see, e.g.,][]{bambi2021}. This includes the BH's spin \citep{reynolds2021}, properties of the disk such as its density and ionization, and, in particular, the location and geometry of the primary source of hard X-rays.  

Earlier studies \citep[e.g.,][]{wilms2001,miller2002,fabian2004,dauser2012,keck2015} have shown that steep disk irradiation profiles, also called emissivity profiles, are required to explain the observations. The simplest % possible
geometry that explains this strong focusing of the irradiation onto the inner accretion disk is the ``lamp post'' geometry \citep[e.g.,][]{martocchia1996,martocchia2002,vaughan2004,dauser2013}, where the primary source geometry is approximated as an emitting point source on the BH's spin axis. The success of this geometry in describing spectra of many sources has been interpreted in a way that the primary source cannot be significantly extended in size \citep[e.g.,][]{fabian2014,parker2015,duro2016,beuchert2017,walton2020,walton2021}, consistent with X-ray timing studies \citep{kara2016,cackett2021} and data from gravitationally lensed quasars \citep[][and therein]{vernardos2024}. 

Several models for relativistic reflection exist to study the physical properties of BHs and their innermost regions \citep[see, e.g.,][]{bambi2024a}. Currently, \relxill \citep{dauser2014, garcia2014} is the most widely used and significantly advanced relativistic reflection model. It predicts the reflected and direct parts of the spectra \citep{dauser2016}, including a self-consistent treatment of the ionization gradient based on the irradiation profile and an $\alpha$-disk model density profile of \citet{shakura1973a} and takes into account returning radiation, % which is 
the secondary reflection produced by reflected photons returning to the disk \citep{dauser2022}. Other advanced models, such as \textsc{reltrans} \citep{ingram2019a,mastroserio2021}, \textsc{kyn} \citep{dovciak2004}, or \textsc{reflkerr} \citep{niedzwiecki2008,niedzwiecki2019} address other aspects such as time-resolved studies, reflection from the plunging region, or polarization. Many of these models also assume a lamp post geometry.

Theoretical and observational work shed light on possible configurations of the innermost region. The primary source can be stationary or outflowing \citep{beloborodov1999,dauser2013} or be variable
% in its radial and vertical size 
\citep[e.g.,][]{wilkins2015,chainakun2017,kara2019}. Several distinct regions are sometimes required to describe the data \citep[e.g.,][]{furst2015,petrucci2018}. Recently, X-ray polarimetry data from several sources have been interpreted as evidence % to support 
for 
a radially extended primary source \citep[e.g.,][]{krawczynski2022}, albeit without fully accounting for effects such as the contribution of % reflected 
returning radiation to the polarization \citep[see][]{ratheesh2024}. 
    
Significant efforts have been made in recent years to describe more realistic primary source geometries. These include vertically extended, line-like primary sources in \relxill \citep{dauser2013} and \textsc{reltrans} \citep{lucchini2023}, radially extended ring-like sources \citep[e.g.,][]{miniutti2003,miniutti2004,suebsuwong2006,niedzwiecki2008}, slab and spherical geometries \citep[e.g.,][]{niedzwiecki2005,wilkins2012,wilkins2015}, and cylindrical and conical geometries \citep[e.g.,][]{gonzalez2017,szanecki2020,uttley2025}. A disk-like geometry was considered by \citet{riaz2022}. More recently, conical and wedge-like geometries have also been implemented in the ray-tracing code \textsc{KerrC} \citep{krawczynski2022a}, while spherical geometry has been included in the ray-tracing code \textsc{Monk} \citep{zhang2019, zhang2024a} and a relativistic reflection model for off-axis point sources was released by \citet{feng2025}.

In this paper, we extend the \relxill framework by allowing the primary source to be radially extended. \relxill is the perfect choice for this generalization, as it is a very fast model allowing for the spectral fitting of complex data sets and includes a detailed description of the accretion disk ionization and density gradient. We chose a ring-like primary source geometry as the simplest extension of the lamp post geometry, since multiple ring sources of different sizes and heights can be readily combined to model extended primary sources of arbitrary axially symmetric geometry. A ring-like model is also a good description of any compact off-axis source, because any off-axis point source will appear as an axially symmetric ring when observed over typical observation timescales of current X-ray observatories. % assuming it orbits along with the disk. 
In Sect.~\ref{sec:model}, we describe the new ring source model. We then summarize in Sect.~\ref{sec:results_model} the influence of the radial size of the primary source on the reflected spectra. In Sect.~\ref{sec:results_data}, we describe how we applied the ring model to the radio-quiet Seyfert~2 galaxy \esosource. Finally, we discuss the new model and the results in Sect.~\ref{sec:general_results}.

% \section{Modeling the reflection from axisymmetric sources} % this is not the specific reflection
\section{Modeling reflection from axisymmetric sources}
    \label{sec:model}
    
\begin{figure}
\centering
\includegraphics[width=\columnwidth]{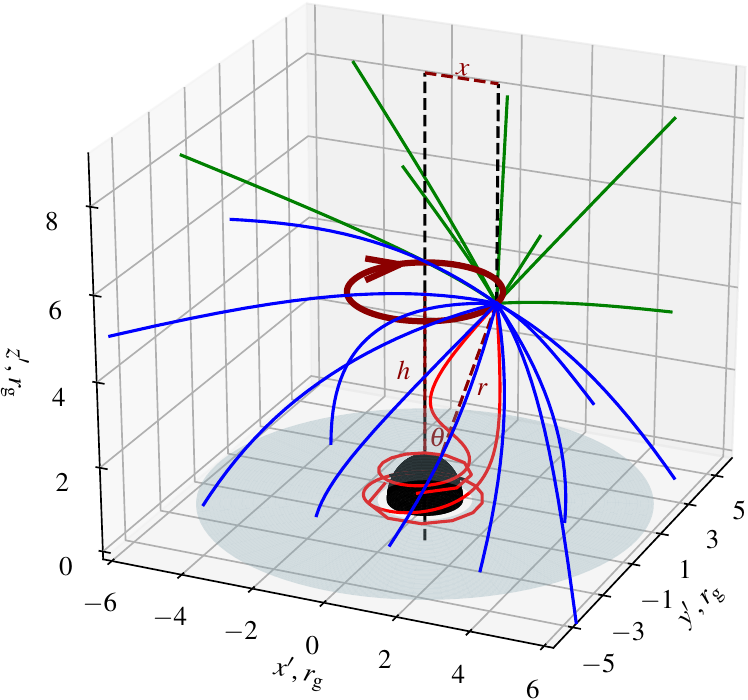}
\caption{Individually selected isotropically distributed photon trajectories emitted by a point source at $h = 5\rg$, $\ringradius = 2\rg$. The BH has a spin of $a=0.998$ and is located at $x^\prime = 0, y^\prime = 0, z^\prime = 0$. The event horizon is illustrated in black. The accretion disk plane is $z^\prime = 0$. The dark red ring is the primary source after axial averaging. Red trajectories plunge into the BH, green trajectories escape the system, and blue trajectories hit the accretion disk, where they will be reflected. The point source is rotating in the direction denoted by the dark red arrow with a velocity $v_{\varphi}$ as given by Eq.~\eqref{eq:velocity}. We also display the corresponding spherical coordinates of the source, $(\sphericalradius, \theta)$, related to $(h, \ringradius)$ by Eq.~\eqref{eq:metric_to_hx}. 
}\label{fig:individual_photon_trajectories}
\end{figure}

We considered a geometrically thin accretion disk irradiated by a primary source, which is described by a geometrically thin ring at height, $h$, above the equatorial plane. The ring has radius, $\ringradius$, and is centered on the BH's rotational axis (see Fig.~\ref{fig:individual_photon_trajectories}), such that $\ringradius=0$ recovers the lamp post geometry. We defined the dimensionless spin of the BH as $a = J / M$, ranging from $-0.998$ to $0.998$ \citep{thorne1974}. Distances are given in units of gravitational radii, $\rg = \mathrm{G} M / \mathrm{c}^{2}$, unless otherwise stated. We set the speed of light, the gravitational constant, and the BH mass equal to unity, $\mathrm{c} = \mathrm{G} = M = 1$. 

\subsection{Basic equations}\label{subsec:basic_equations}
    
For our ray tracing procedure, we used the \citet{kerr1963} metric in \citet{boyer1967} coordinates \citep{bardeen1972}, % expressed as 
\begin{equation}
\label{eq:kerr_metric}
    \dif s^2 = - e^{2\nu} \dif t^2 + e^{2 \psi} \left(\dif \varphi - \omega \dif t \right)^2 + e^{2\mu_1} \dif \sphericalradius^2 + e^{2 \mu_2} \dif \theta^2 \;,
\end{equation}
where    \begin{multline}
    \label{eq:coefficients_metric}
    e^{2\nu} = \frac{\Sigma \Delta}{A}, \hspace*{0.5em} e^{2\psi} = \frac{A \sin^2\theta}{\Sigma}, \hspace*{0.5em}
    e^{2\mu_1} = \frac{\Sigma}{\Delta}, \hspace*{0.5em} e^{2\mu_2} = \Sigma, \hspace*{0.5em} \omega = \frac{2 a \sphericalradius}{A} \;, 
\end{multline}
with    \begin{multline}\label{eq:notations_metric}
    \Delta = \sphericalradius^2 - 2 \sphericalradius + a^2, \hspace*{0.5em} \Sigma = \sphericalradius^2 + a^2 \cos^2\theta, \hspace*{0.5em}
    A = (\sphericalradius^2 + a^2)^2 - a^2 \Delta \sin^2\theta \;.
\end{multline} 
and where the source height, $h$, and radius, $\ringradius$, are given by
\begin{equation}\label{eq:metric_to_hx}
    h = \sphericalradius \cos\theta, \hspace*{0.5em} \ringradius = \sqrt{\sphericalradius^2 + a^2} \sin\theta. \;
\end{equation}    
The inner edge of the geometrically thin accretion disk in the equatorial plane around the BH is given by the innermost stable circular orbit (ISCO), \risco. The Keplerian velocity of the accretion disk is \citep{bardeen1972}:
\begin{equation}
    \OmegaKeplerian = \left(\diskradius^{3/2} + \left|a\right|\right)^{-1} \;,
\end{equation}
where \diskradius is the spherical radius, \sphericalradius, measured in the metric equatorial plane, namely, the disk radius. 

The Lorentz-factor, in general and for circular equatorial orbits \citep{bardeen1972}, is
\begin{equation}\label{eq:Lorentz_factor_disk}
    \gamma = \left(1 - v_\varphi^2\right)^{-1/2} = \frac{\sqrt{\diskradius^2 - 2 \diskradius + a^2} \left(\diskradius^{3/2} + \left|a\right|\right)}{\sqrt{\diskradius^{2} - 3 \diskradius + 2 a\diskradius^{1/2}} \sqrt{\diskradius^3 + \diskradius a^2 + 2 a^2}}\;.
\end{equation}
We use the common assumption \citep[e.g.,][]{miniutti2004,niedzwiecki2005,niedzwiecki2008,wilkins2012} that the primary source is corotating with the accretion disk. This assumption is more plausible than a stationary source if we are positing that the primary source consists of material ejected from the disk and some kind of (partial) angular momentum conservation holds. For rings outside the ISCO, $\ringradius \geq \risco$, the angular velocity of the source is the Keplerian velocity of the accretion disk at distance, $\diskradius = \ringradius$. Following \citet{niedzwiecki2008}, inside the ISCO, $\ringradius < \risco$, the angular velocity of the ring, \OmegaSource, is assumed to be constant and given by the disk velocity at \risco. The only nonzero component of the three-velocity is, therefore, 
\begin{equation}\label{eq:velocity}
    v_{\varphi} = e^{\psi - \nu} \left(\OmegaSource - \omega\right),
\hspace*{0.5em} \mbox{with} \hspace*{0.5em}
    \OmegaSource = \left(\max(\risco, \ringradius)^{3/2} + \left|a\right| \right)^{-1}\;.
\end{equation}
For $\ringradius = 0$, $v_{\varphi} = 0$, meaning a stationary point source. However, we stress that the actual velocity in the plunging region inside the ISCO is more complicated, as radial and vertical components of the velocity become non-negligible \citep[see, e.g.,][]{pu2016, mummery2023}. Thus, the photons that are beamed in the direction of motion will redistribute the resulting disk irradiation \citep[see, e.g.,][]{niedzwiecki2008,dauser2013}. We also note that the \relxill model allows for a vertical velocity of the primary source and, therefore, it includes the beaming of the flux in the direction of the rotation axis. 
Contrary to the case of a corotating source, a static source ($v_{\varphi} = 0$, but note frame dragging, $\omega$) will boost less photons to the outer disk and there will be less of a Doppler shift, such that the blueshift in the line decreases \citep[see, e.g.,][for a comparison of static and corotating ring sources]{niedzwiecki2008}. 

With these assumptions, following \citet{niedzwiecki2005} the energy shift of the photons from a source with initial rest frame energy, \Esource, to the energy, \Edisk, at the point of impact with the disk is    \begin{equation}\label{eq:energy_shift}
    g = \frac{\Edisk}{\Esource} = \frac{\left[\gamma e^{-\nu} \left(1 - \OmegaKeplerian (\diskradius) \lambda\right)\strut\right]_\mathrm{disk}}{\left[\gamma e^{-\nu} \left(1 - \OmegaSource (\ringradius) \lambda\right)\strut\right]_\mathrm{source}},
\end{equation}
where $\gamma$, $e^{-\nu}$, \OmegaSource, and \OmegaKeplerian are taken at the locations of the source (``source'') and the disk (``disk''), respectively, $\lambda = L/E$ is the dimensionless angular momentum of the photon\footnote{$L$ is the angular momentum component of the photon parallel to the BH rotational axis and $E$ is the total energy \citep[see][for the details of definition and angular dependence]{bardeen1972,niedzwiecki2005}.} (see Appendix~\ref{app:energyshift} for the derivation of this equation). Equation~\eqref{eq:energy_shift} reduces to the lamp post case for $\lambda=0$ \citep[see, e.g.,][]{dauser2013}. For a given $(\sphericalradius, \theta, \varphi)$, $\lambda$ depends on the initial direction of motion of the photon and can take positive and negative values.

\subsection{Ray-tracing for an off-axis source}
\label{subsec:raytracing}

To perform ray-tracing, we use the \ynogk code for null-geodesics in the Kerr metric \citep{yang2013}, computing the velocity profile of the emitting source with Eq.~\eqref{eq:velocity}. For each parameter combination, we use a Monte Carlo approach. We assume a point source that emits isotropically in its rest frame, drawing the emission angles of individual photons from uniform distributions for $\varphi \in [0, 2\pi)$, and $\cos\theta \in [-1, 1]$.

Figure~\ref{fig:individual_photon_trajectories} shows selected simulated photon trajectories for a rapidly rotating BH with a spin of $a=0.998$. In the frame of rest of an observer at infinity, the photons are boosted in the direction of the rotation due to the fast azimuthal velocity of the source. Depending on their initial direction, photons emitted from the ring will either fall into the BH, hit the accretion disk, or escape to infinity. We assume that photons crossing the equatorial plane between the ISCO and the event horizon are scattered or advected inward by the in-flowing matter \citep{agol2000} and are therefore captured by the BH. 

\subsection{Disk irradiation by a ring source}
\label{subsec:disk_irradiation}
 
A ring source can be modeled as a combination of multiple off-axis point sources at different azimuthal angles. Due to the azimuthal symmetry, each of these sources will produce the same irradiating flux on the disk, such that the irradiation profile (i.e., the radial-dependent incident flux on the accretion disk) can be derived by simulating a single off-axis point source and then integrating the resulting flux distribution on the accretion disk over the azimuthal coordinate \citep[see also][]{wilkins2012}.

We performed a set of simulations for different values of spin, $a$, height, $h$, and source radius, \ringradius, at a fixed azimuthal angle $\phi = 0$. From these simulations, we derived the specific photon flux, $\fluxspecific(\diskradius, \phi, \Esource)$, incident on a disk element at radius, \diskradius, and azimuth, $\phi$, of width, $\dif \diskradius$ and $\dif \phi$, and an energy, $\Esource$, in the source rest frame. The flux is proportional to the number of photons incident on a given disk area divided by this area. Following \citet{dauser2013}, for each disk annulus, we determined the irradiating flux at energy $\Edisk = g \Esource$ (Eq.~\ref{eq:energy_shift}) in the frame of rest of the disk by integrating over the azimuth and radius,
\begin{equation}\label{eq:irradiating_flux}
  F (\diskradius, h, \ringradius, \Edisk) = \frac{1}{\Kerrarea \gamma} 
  \int\limits_{0}^{2\pi} \int\limits_{\diskradius - \Delta \diskradius / 2}^{\diskradius + \Delta \diskradius / 2} \fluxspecific \left(\diskradius^\prime, \phi^\prime, \frac{\Edisk}{g}\right) \diskradius^\prime \dif \diskradius^\prime \dif \phi^\prime \;,
\end{equation}
where $\Kerrarea$ is the area of the annulus at radius, \diskradius, and width, $\Delta \diskradius$, as seen by a distant observer % , given by 
\citep{wilkins2012}:
\begin{equation}\label{eq:disk_area}
    {\Kerrarea} (\diskradius, \Delta \diskradius, a) = 2 \pi \Delta \diskradius \sqrt{\frac{\diskradius^4 + a^2 \diskradius^2 + 2 a^2 \diskradius}{\diskradius^2 - 2 \diskradius + a^2}} \;, 
\end{equation}
and $\gamma$ is the Lorentz factor of the disk element (Eq.~\ref{eq:Lorentz_factor_disk}), accounting for the disk's rotation. The proper area of the disk annulus in the frame of the disk is therefore given by $\gamma \Kerrarea$.

As the irradiating flux depends on the primary source spectral shape, integration over the azimuthal coordinate in Eq.~\eqref{eq:irradiating_flux} can be performed only for a known spectral shape. Assuming that the primary source emits a power-law spectrum in its rest frame\footnote{In all \relxill models with a primary \nthcomp spectrum, we use the approximation that the spectrum follows a power law in the relevant energy range, therefore the same transformation applies.} \citep[see][]{dauser2013}, $\fluxspecific(\Esource) \propto \Esource^{-\Gamma}$, with photon index $\Gamma$, and using discrete annuli with sufficiently small $\Delta \diskradius$,
\begin{equation}\label{eq:irradiating_flux_powerlaw}
    F (\diskradius, h, \ringradius, \Edisk) = \frac{2\pi \diskradius \Delta \diskradius}{\Kerrarea\gamma} \sum_{\phi_i} \fluxspecific\left(\diskradius, \phi_i, \Edisk\right) g^{\Gamma}(\diskradius, \phi_i) \;.
\end{equation}

\subsubsection{Energy shift from the off-axis sources}
\label{subsec:energy_shift_spread}

The main difference between a ring and a lamp post geometry is that there is a large range of different photon trajectories incident on the same disk radius. This is because photons emitted from different locations on the ring can hit the same disk element. Since the Doppler boosting can be strongly different for these photons, this affects the irradiation profile seen by the accretion disk.

\begin{figure}
    \centering
    \includegraphics[width=\columnwidth]{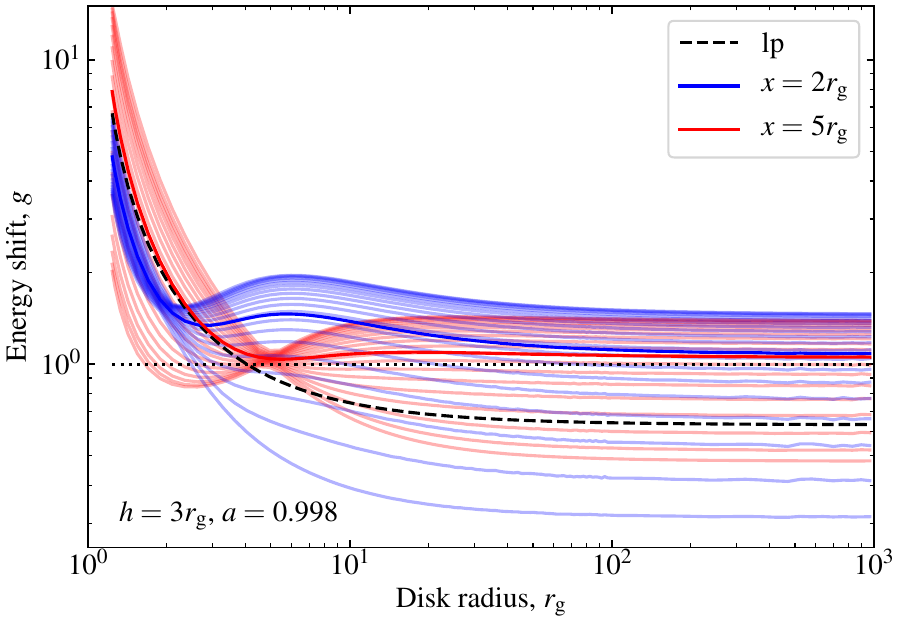}
    \caption{Energy shift, $g$, of  photons propagating from off-axis sources, located at $h = 3\rg$, at two ring radii $\ringradius = 2\rg$ (blue lines) and $\ringradius = 5\rg$ (red lines), and BH spin of  $a = 0.998$. Photon trajectories are selected from an isotropic distribution of the initial photon directions in the rest frame of the sources. Lighter lines show energy shifts of photons emitted by the off-axis sources located on the same ring at different azimuthal angles. Thick lines represent the resulting averaged energy shifts over the disk annuli. The  dashed black line corresponds to the energy shift of the photons from a lamp post source with $h = 3\rg$. 
}\label{fig:energy_shift}
\end{figure}
In Fig.~\ref{fig:energy_shift}, we show how the energy shift of isotropically emitted photons from off-axis sources located on two compact rings at $h=3\rg$ depends on radius. For a very compact ring, $\ringradius=2\rg$, all photons emitted by different points on the ring experience a similar energy shift at small disk radii, increasing to factors of $\sim$20 at large radii. An increase in ring radius ($\ringradius=5\rg$) leads to a larger spread of energy shifts at small radii, and a lower spread at large radii, respectively. As the ring is corotating with the disk, for $\diskradius \sim \ringradius$  the range of energy shifts drops to zero. For radii larger (smaller) than the corotation radius, the disk rotation is slower (faster) than that of the emitting source. We note that for a given $\diskradius$ the time-dilation term of the metric (Eq.~\ref{eq:kerr_metric}), $e^{-\nu}$, also influences the energy shift, causing it to decrease. 

For a qualitative comparison with the lamp post, we also computed the average energy shift of all photons incident on the given disk annulus (Fig.~\ref{fig:energy_shift}, thick lines). Averaging this value over azimuth yields the energy shift of the ring. The energy shift is a blue-shift, $g > 1$, for all radii, leading to a Doppler boost of the irradiating flux. In contrast, irradiating photons from a lamp post are redshifted at large radii, resulting in a reduced flux (Fig.~\ref{fig:energy_shift}, dashed line). For sources that are similar to a lamp post ($\ringradius \lesssim \risco$) and close enough to the BH ($h$ less than a few \rg), the emitted photons will also be red-shifted if incident at large radii.

\subsubsection{Disk irradiation profiles}
\label{subsec:disk_irradiation_profiles}

\begin{figure}
    \centering
    \includegraphics[width=\columnwidth]{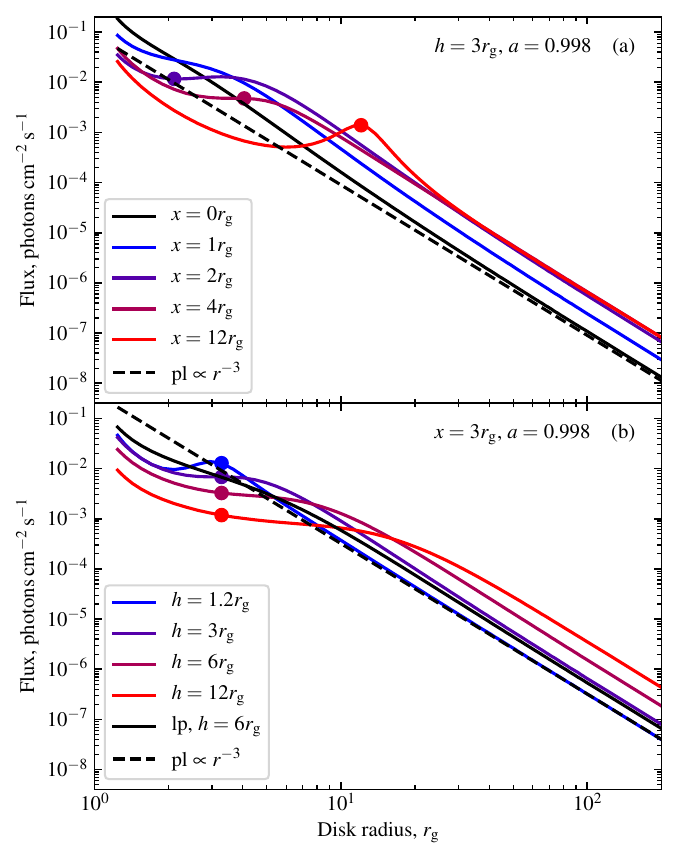}
    \caption{Disk irradiation flux profiles for varying emitting ring radii, \ringradius, for (a) a  ring at $h = 3\rg$ and (b) a ring of varying height with radius fixed at $\ringradius = 3\rg$, a BH spin of $a = 0.998$, and an irradiating source with $\Gamma = 2$. The solid black line is for the lamp post case, $\ringradius = 0$, for $h=3\rg$ in (a) and $h=6\rg$ in (b). The  dashed line shows a power-law flux $\propto \diskradius^{-3}$. Markers of the same color on each curve show the disk radius equal to the primary source radius, $\diskradius \equiv \ringradius$. The normalization of the fluxes is set to the non-relativistic limit at large radii (see Appendix~\ref{app:implementation}). }\label{fig:flux_disk}
\end{figure}

Using Eq.~\eqref{eq:irradiating_flux}, we calculated the irradiating flux on the accretion disk, which is often called the ``emissivity profile,'' for a ring-like primary source. Figure~\ref{fig:flux_disk} shows a selection of characteristic
irradiating profiles for rings at different heights and radii. In general, the profiles follow the typical steepening toward
the ISCO and converge toward the flat-space irradiating flux of $F \propto \diskradius^{-3}$ for large radii. The ring geometry
also leads to a flattening of the profile at radii directly underneath the ring.

For a fixed source height, with increasing source radius the flux is redistributed towards larger disk radii (Fig.~\ref{fig:flux_disk}a). For $\ringradius \gtrsim h$, a secondary peak forms below the ring. This behavior is expected for simple geometric reasons. For a source relatively close to the disk, $h \lesssim \ringradius$, most photons will hit the disk below the ring, forming a local flux maximum at $\diskradius \sim \ringradius$, if the source is more extended radially than vertically. In addition, light-bending modifies the irradiating profiles (Sect.~\ref{subsec:raytracing}), as radiation is focused onto the inner part of the disk. Similar to the lamp post geometry, this effect is strongest for $\diskradius < 10\rg$, and further increases irradiation at the disk radius directly underneath the ring, $\diskradius \approx \ringradius$. For the smallest ring radii, the local flux maximum emerges at slightly larger radii than the ring radius because light bending ``pulls'' the flux emitted directly below the ring away to even smaller radii.

For a fixed ring radius, \ringradius, with increasing height more flux impacts the outer disk and less flux is focused towards the inner edge (Fig.~\ref{fig:flux_disk}b). Even for the highest ring source considered here, $h = 12\rg$, the irradiation profile is still steep at the innermost radii. As long as $\ringradius \lesssim h$, the irradiation profile stays similar to that of a lamp post at a similar or larger height. The local maximum at $\diskradius \sim \ringradius$ is replaced by a flattening for rings with $h > \ringradius$. Crucially for the physical interpretation of observed data, we find that the irradiation profiles due to ring sources with relatively small radii can mimic the irradiation profiles of lamp posts with larger heights. For example, profiles with $h = 3\rg$, $\ringradius = 3\rg$ and $h = 6\rg$, $\ringradius = 0\rg$ differ significantly only over a very small range of radii. 

\section{Ring source model for relativistic reflection}
\label{sec:results_model}

With the irradiation profiles for the extended ring-like primary source, we calculated the relativistic line broadening and the resulting reflection spectra for such a geometry. We implemented them as part of the \relxill model framework \citep{dauser2014,garcia2014}, such that the model can directly fit the observed spectra of accreting BH systems and constrain the parameters of the primary source. We refer to Appendix~\ref{app:comparison} for a verification of our code.

\subsection{Implementation of the ring geometry in \relxill}
    \label{subsec:implementation}

The \relxill framework is well suited to implement a ring-like primary source, as it is designed in a modular way and we only need to implement a new irradiation profile (Eq.~\ref{eq:irradiating_flux}). The major complexity is to account for the large spread of energy shifts (Sect.~\ref{subsec:energy_shift_spread}) and to apply the correct normalization between direct and reflected flux, including a relativistic lensing of the ring source \citep[see also][]{ingram2019a,feng2025}. After that we re-use the full \relxill setup for the computation of the non-relativistic reflection in the rest-frame of the accretion disk \citep{garcia2010, garcia2013} and its propagation to the observer \citep{dauser2010}.
    
In \relxill, the disk irradiation profile is constructed from pre-calculated tables that contain the relevant parameters of the ray-tracing simulations. For a given set of input parameters, the table values are linearly interpolated. To implement the ring source in \relxill, we therefore pre-calculate a table extending the spin-height grid to include the new radius parameter, \ringradius. To obtain a well-sampled photon flux, $\fluxspecific$, we simulate $10^8$ isotropically distributed photons for each grid point. Additionally, the energy shift values, $g$, are stored, such that the irradiation profile can be calculated for any value of the photon index $\Gamma$ from Eq.~\eqref{eq:irradiating_flux_powerlaw} or any more general spectral shape, Eq.~\eqref{eq:irradiating_flux}. As there is a range of energy shifts of photons incident on one annulus at radius, $\diskradius$ (Sect.~\ref{subsec:energy_shift_spread}), we bin the photon trajectories in energy shift in 20 bins and store the relative flux value of each bin in the table (see Appendix~\ref{app:implementation}). 
In order to calculate the primary spectrum normalization and reflection fraction, we also store the photon fractions reaching the disk, \fractiondisk, infinity, \fractionesc, and BH, \fractionblackhole. As the direct spectrum is also affected by the energy shift and gravitational lensing, we also store these data (see Appendix~\ref{app:energyshift} for the detailed calculation).

By adding the radial size of the ring source, $\ringradius$, we were able to define a new flavor of the \relxill model, named \relxilllpext. We built it on the basis of the most advanced \relxill model, \relxilllpcp, which assumes a Comptonization primary spectrum modeled by \nthcomp \citep{zdziarski1996}. We also include effects of the self-consistent ionization gradient in the disk and a self-consistent treatment of the returning reflected radiation, based on \citep{dauser2022}. The model is publicly available as part of the \relxill model package\footnote{\href{https://www.sternwarte.uni-erlangen.de/research/relxill/}{https://www.sternwarte.uni-erlangen.de/research/relxill/}}.

\subsection{Line reflection spectra}
\label{sec:results_model_iron}

\begin{figure*}
  \centering
  \includegraphics[width=\textwidth]{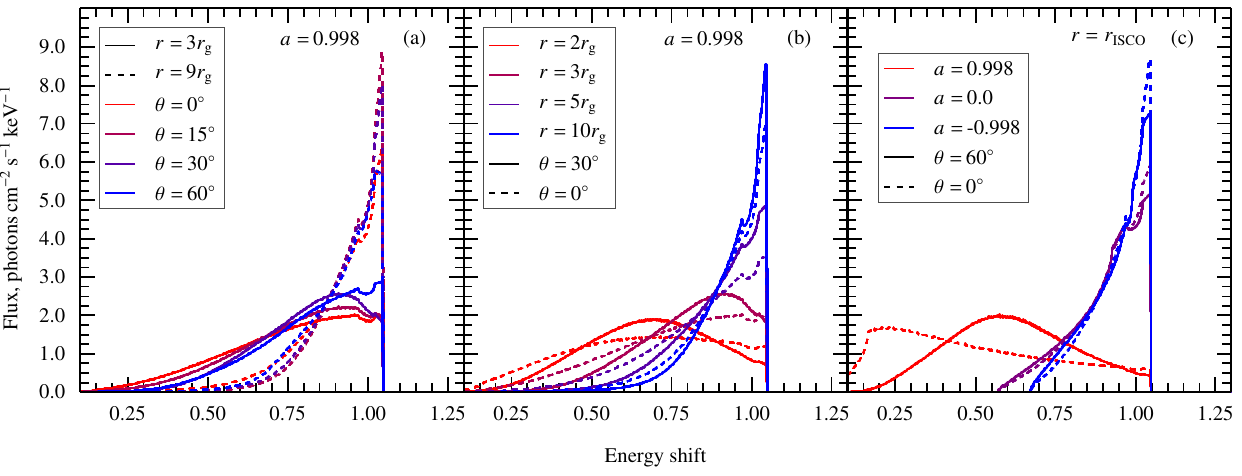}
  \caption{Line profiles for a ring geometry simulated with \relxill. We display the geometric parameters of the ring in terms of spherical radii, \sphericalradius, and polar angles, $\theta$. The parameters are uniquely related to height and ring radius  through Eq.~\eqref{eq:metric_to_hx}. For ease of comparison, we also convert these values to heights and radii in Table~\ref{table:rtheta2hx}. (a) Varying polar angle of the primary source, $\theta$, for two spherical radii, $\sphericalradius = 3\rg$ (dashed lines) and $\sphericalradius = 9\rg$ (solid lines), with $a = 0.998$. (b) Varying the spherical radius of the primary source, \sphericalradius, with two polar angles of the primary source, $\theta = 30^{\circ}$ (solid lines) and $\theta = 0^{\circ}$ (dashed lines, the lamp post case), with $a = 0.998$. (c) Varying the BH spin, $a$, for a source spherical radius fixed at ISCO value, $\sphericalradius = \risco$, and $\theta = 60^{\circ}$. ISCO radii are $1.24\rg$, $6\rg$, $9\rg$ for $a = 0.998$, $0.0$, $-0.998$, respectively. To convert the quantities $(\sphericalradius, \theta)$ to $(h, \ringradius)$, we provide the Table~\ref{table:rtheta2hx}. All lines are normalized to have the same integrated photon flux. The photon index is $\Gamma = 2$ and the inclination to the observer is $i = 30^{\circ}$.}\label{fig:relline_spectra}
\end{figure*}

We investigated the effect of the location and radius of the ring-like primary source on a single relativistic line profile. As general line profiles of radially extended sources have been presented in a variety of publications \citep[e.g.,][]{miniutti2003,miniutti2004,niedzwiecki2008,dauser2010,dauser2013,riaz2022}, here we focus on comparing the ring sources with their most relevant counterparts, point-like lamp post sources. For the comparison, we used spherical coordinates $(\sphericalradius, \theta)$ to define the source position (Eq.~\ref{eq:metric_to_hx}), since \sphericalradius has the most significant effect on the line shape and the angular position of the source has only a secondary effect due to Doppler shifts (as we explain  later in this section). For 
clarity, conversions between the quantities $(\sphericalradius,\theta)$ and $(h, \ringradius)$ are given in Table~\ref{table:rtheta2hx}. A lamp post source at height $h = \sphericalradius$  corresponds to $\theta = 0^{\circ}$. 

Figure~\ref{fig:relline_spectra} shows line profiles for sources at different spherical radii and polar angles. For any source with a distance to the BH of $\sphericalradius = 3 \rg$ (Fig.~\ref{fig:relline_spectra}a), the line is very broad, with a strong red wing. For increasing polar angle of the source (i.e., increasing ring size), the velocity of the ring increases from zero to its maximum value at the ISCO, such that more photons are boosted away from the inner disk and experience a larger blueshift. This increase in velocity compared to the stationary lamp post causes the blue wing of the line to slightly brighten and the overall line shape to narrow. 

Increasing the source distance to the BH (i.e., expanding the spherical radius) leads to the line narrowing. This effect prevails regardless of the polar angle (i.e., the ring radius). The main reason for this behavior is that less flux is focused towards the BH and, therefore, fewer reflected photons come from the innermost accretion disk. For $\sphericalradius \gtrsim 5 \rg$, a double-horned skew-symmetric line feature arises from the strong Doppler shift of the photons due to both the disk and the source rotation.

Figure~\ref{fig:relline_spectra}c shows how the line shape changes with spin, $a$, for a source located at a distance to the BH equal to the inner edge of the accretion disk, that is, $\sphericalradius = \risco$. As the inner edge of the accretion disk moves outwards for a smaller BH spin, this distance \sphericalradius also increases. We show sources with two polar angles, a lamp post with $\theta= 0^{\circ}$, and a source with $\theta= 60^{\circ}$, corresponding to a source almost above the ISCO, at a height of $0.5\risco$ (see Table~\ref{table:rtheta2hx}). As commonly known \citep[see, e.g.,][]{dauser2013}, for any disk irradiation a broad line is produced only in the case of high spin, which is also the case for our ring-like source \citep[however, see][for a discussion of emission from the plunging region]{reynolds1997}. Differences between the line profiles for a lamp post and the ring source are more distinct in the high spin case. For non-rotating and counter-rotating BHs, the line shapes are more similar, irrespective of the angle. 

In summary, a ring primary source instead of a lamp post source does not lead to substantially different line profiles. The most important result is that the line profiles depend mainly on the distance to the BH (i.e., the spherical radius). Increasing the polar angle, $\theta$, leads to a slightly narrower line that is similar to slightly increasing the height of the lamp post. This can for example be seen in Fig.~\ref{fig:relline_spectra}b, where the line shape of the lamp post for $(\sphericalradius=10\rg, \theta=0^\circ)$ is in between the ring source for $(\sphericalradius=5\rg, \theta=30^\circ)$ and $(\sphericalradius=10\rg, \theta=30^\circ)$.  

\begin{table}
\caption{Values of $(\sphericalradius, \theta)$ and corresponding $(h, \ringradius)$ for several spins.}\label{table:rtheta2hx}
\centering
\begin{tabular}{c|c c c c c c c}
    \hline
    \hline
    Values & \multicolumn{7}{c}{Plot (a)} \\
    \hline
    $a$ & 0.998 & -- & -- & -- & -- & -- \\
    $\sphericalradius$, \rg & 3 & 3 & 3 & 9 & 9 & 9 \\
    $\theta$, $^{\circ}$ & 15 & 30 & 60 & 15 & 30 & 60 \\
    $h$, \rg & 2.9 & 2.6 & 1.5 & 8.7 & 7.8 & 4.5 \\
    \ringradius, \rg & 0.8 & 1.6 & 2.7 & 2.3 & 4.5 & 7.8 & \\
    \hline
    Values & \multicolumn{4}{c}{Plot (b)} & \multicolumn{3}{c}{Plot (c)} \\
    \hline
    $a$ & 0.998 & -- & -- & -- & -- & 0 & $-0.998$ \\
    $\sphericalradius$, \rg & 2 & 3 & 5 & 10 & 1.24 & 6 & 9 \\
    $\theta$, $^{\circ}$ & 30 & 30 & 30 & 30 & 60 & 60 & 60 \\
    $h$, \rg & 1.7 & 2.6 & 4.3 & 8.7 & 0.6 & 3.0 & 4.5 \\
    \ringradius, \rg & 1.1 & 1.6 & 2.5 & 5.0 & 1.4 & 5.2 & 7.8 \\
    \hline\hline
\end{tabular}
\tablefoot{The table shows the values in spherical coordinates, displayed in Fig.~\ref{fig:relline_spectra} and corresponding values in terms of height and ring radius, obtained with Eq.~\eqref{eq:metric_to_hx}. We do not display cases where $\theta = 0^\circ$,  because, in these cases, $\sphericalradius \equiv h$.}
\end{table}

\subsection{Relativistic reflection spectra}
\label{sec:results_model_continuum}

\begin{figure}
  \centering
  \includegraphics[width=\columnwidth]{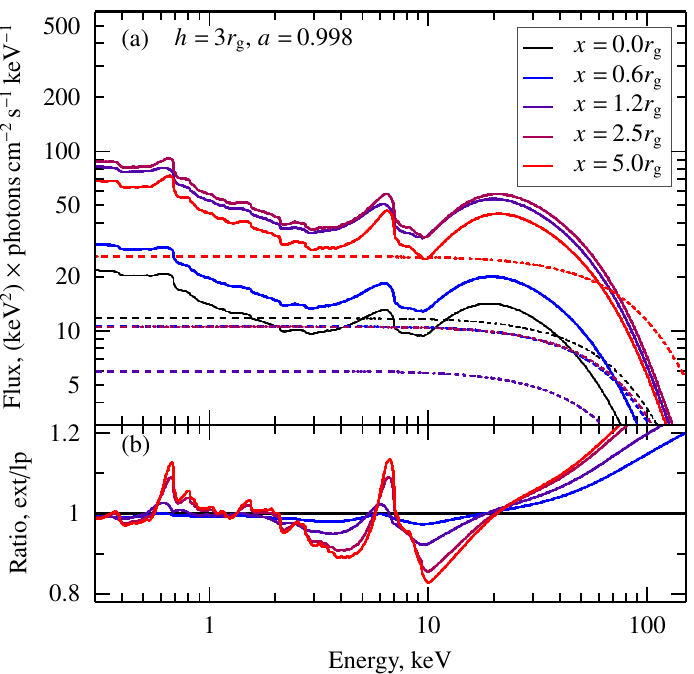}
  \caption{Relativistic reflection spectra for varying ring radius, $\ringradius$. (a) Solid lines show simulated reflected spectra for BH spin of $a = 0.998$, height of the primary source, $h = 3\rg$, and varying ring radius. Dashed lines show the primary spectra. Different colored lines correspond to different ring radii. The black line corresponds to the lamp post case, $\ringradius = 0$. The spectra are normalized setting $\boost = 1$ in \relxill, such that the reflection fraction is predicted from the geometry of the isotropically emitting primary source. Returning radiation is turned off. All other parameters of \relxill are set to their default values, i.e., the iron abundance is $\feabund = 1$, the inclination is $i = 30^{\circ}$, the disk inner edge is at $\risco$, the disk outer edge is at $400\rg$, the photon index is $\Gamma = 2$, the disk ionization is $\logxi = 3.1$, the disk density $\logdensity = 10^{15}\,\cmminusthree$, and the electron temperature $\kTe=60$\,keV. (b) Ratio between the given reflected spectrum of the ring and the lamp post spectrum, re-normalized to the same flux.
}\label{fig:relxill_refl_spectra}
\end{figure}

To illustrate the effects of changing normalization and energy shifts on the spectra, we show (in Fig.~\ref{fig:relxill_refl_spectra}) both the direct and reflected components of the relativistic reflection spectra for ring-like primary sources. We assumed a maximally spinning BH, $a = 0.998$, and a source at a small height above the disk plane, $h=3\rg$, in this case and disregard the returning radiation to focus on the main effect of changing the radius of the ring-like source. We keep all other \relxill parameters at their default values. 

We find that the total observed flux increases monotonically with increasing ring radius. The primary component drops for $\ringradius \sim \risco \approx 1.2 \rg$, while the reflected component increases rapidly with $\ringradius$. This behavior is due to a combination of geometric and relativistic effects. The reflected flux increases strongly with the energy boost from the source to the disk, with fewer photons advected into the BH. The fraction of photons hitting the disk reaches its maximum at $\sim 1.8 \rg \approx 1.5\risco$, while the escaping fraction has a minimum at the same radius and, hence, the reflection fraction also reaches its maximum. For small ring sizes, the primary flux drops due to general-relativistic lensing near the BH, which redistributes more flux towards high inclinations with respect to the observer, that is, even if the flux escapes, it still bends toward the equatorial plane and disk. For inclinations of $i \gtrsim 60^{\circ}$, the lensing focuses the flux toward the observer, thus increasing the primary component. 

For sources with $\ringradius\sim 1\text{--}2\risco$ ($1.2$--$2.5$\rg), the equivalent width of the broad iron line reaches its maximum, which is about three\ times higher than that found for lamp posts. As noted above, the primary continuum reaches its minimum level for the same ring radii. For larger source radii, the equivalent width stays above the lamp post level, although lower than the ISCO case. This increase in the equivalent width (i.e., stronger lines) found for ring sources can potentially influence the measurement of iron abundances in spectral fits; in other words, fitting a lamp post would yield a higher iron abundance to explain a broad line than models with an extended source. Overall, a source above the ISCO produces the strongest relativistic reflection relative to the primary flux and, therefore, also the strongest broad lines.

\section{Application to ESO\,033-G002}
\label{sec:results_data}

To illustrate the application of the new reflection model, we went on to apply it to an observation of the Seyfert-2 active galactic nucleus (AGN) \esosource that is well known to show strong relativistic reflection. The data have been analyzed in detail by \citep[][hereafter \Waltontwentyone]{walton2021}. These authors showed that this source has a strong relativistic reflection component in the spectrum, which can be described well by a lamp post primary source very close to the BH ($h \lesssim 2 \rg$). The strong reflection in this source makes it a well-suited target for a first application of our new model. 

\subsection{Observation and data reduction}
\label{subsec:Data}

We use data from a joint observation of \esosource in 2021 June with \xmm \citep[][ObsID 0863050201, 109\,ks for EPIC-pn and 125\,ks for both EPIC-MOS detectors]{jansen2001} and \nustar \citep[][ObsID 0863050201, 172\,ks]{harrison2013}, which were published by \Waltontwentyone.

For \xmm's instruments EPIC-pn, EPIC-MOS1, and EPIC-MOS2, we reduced and extracted the data over the 0.3--10\,keV energy band with \xmm SAS version~21.0.0. For the extraction, we used a circular region with a $35''$ radius; for the background, we used a circular region with $70''$ radius on the same chip. We set \textit{applyabsfluxcorr=yes} to obtain a better agreement between the effective areas of the EPIC-detectors and \nustar, which led to larger cross-calibration constants between \xmm and \nustar compared to \Waltontwentyone\footnote{See \href{https://www.cosmos.esa.int/web/xmm-newton/ccf-release-notes}{https://www.cosmos.esa.int/web/xmm-newton/ccf-release-notes} and file XMM-CCF-REL-388 about the empirical correction of the EPIC effective area.}. After extraction, data from EPIC-MOS1 and EPIC-MOS2 are combined into one dataset. We use optimal binning \citep{kaastra2016} to rebin the spectra, ensuring a minimum signal-to-noise ratio of 5. 

We reduce and extract 4--78\,keV spectra for \nustar-FPMA and -FPMB with the \nustar Data Analysis Software version 2.1.4 and \nustar calibration database v20241015. We extracted the source products and response from a circular region with $90''$ radius and estimated the background from a $100''$ radius region on the same chip. The spectra were again binned with the optimum binning algorithm, ensuring a signal-to-noise ratio of 5. We note that the \nustar spectrum becomes background-dominated at ${\gtrsim} 40$\,keV, resulting in only a few wide bins. 

\subsection{General model set-up}
\label{subsec:ModeltoData}

To fit the data, we used ISIS \citep{houck2000}, applying a similar model as \Waltontwentyone. We used the most recent \relxill model (v2.7) for the lamp post (\relxilllpcp) and the ring source geometry (\relxilllpext) described in Sect.~\ref{subsec:implementation}. Following \Waltontwentyone, we assumed that the relativistic reflector is ionized, modeled by \xstar with the same model grid as \Waltontwentyone and neutral absorption via the full and partially covering absorbers, $\tbabs_\mathrm{full}$ and $\tbabs_\mathrm{part}$, respectively \citep{wilms2000}. We used the \tbnewfeo model, which has the iron abundance of the absorber as a free parameter, tied to the \relxill iron abundance. We model the distant reflection with \xillvercp and the scattered nuclear flux with \nthcompscat. The \textsc{cp} flavor of the \relxill model also includes a separate \nthcomp component that represents the primary continuum. Different to \Waltontwentyone we describe the emission spectrum from collisionally ionized diffuse gas with \apec\footnote{\href{http://atomdb.org/}{http://atomdb.org/}} \citep{smith2001,foster2012} instead of \textsc{mekal} \citep{mewe1986,liedahl1995}. All of the model components are cosmologically redshifted by the host galaxy redshift $z = 0.0181$ \citep{tueller2010} and affected by Galactic absorption, which we modeled with $\tbabs_\mathrm{Gal}$ for a fixed absorption column of $\NH = 8.95 \times 10^{20}\,\cmminustwo$ \citep{hi4picollaboration2016}. We used the \texttt{wilm}-abundances \citep{wilms2000}, instead of \texttt{grsa} \citep{grevesse1998} used by \Waltontwentyone. Since we applied a flux correction to the data, as recommended by the new extraction guides (Sect.~\ref{subsec:Data}), we took into account the cross-calibration between \nustar and EPIC-pn by applying a multiplicative constant \detconst to the \nustar spectra. 

The full expression of the base model is: $\detconst \times \tbabs_\mathrm{Gal}$ $\times$ (\apec + \nthcompscat + \xillvercp + ($\tbabs_\mathrm{full}$ $\times$ $\tbabs_\mathrm{part}$ $\times$ \xstar $\times$ \relxill)). \relxill stands for the relativistic reflection model, which is either the lamp post model \relxilllpcp in the base model or the new extended model from this work. 

\subsection{Lamp post models}\label{subsec:results_lp_fits}

\begin{table}
\caption{Comparison of the lamp post and extended models for \esosource.}\label{table:parameters_uncertainties_ext}
\centering
\begingroup
\small
\renewcommand{\arraystretch}{1.3}
\begin{tabular}{@{}cccccc@{}}
    \hline\hline
    Parameter & \modellpold & \modellp & \modelextfreeboost & \modelextfixboost\\ 
    \hline
    \multicolumn{5}{l}{$\tbabs_\mathrm{full}$}\\
    \NH & $1.65^{+0.18}_{-0.20}$ & $1.59^{+0.20}_{-0.31}$ & $1.56^{+0.22}_{-0.36}$ & $1.55^{+0.23}_{-0.30}$\\ 
    \hline
    \multicolumn{5}{l}{$\tbabs_\mathrm{part}$}\\ 
    \NH & $7.8\pm1.3$ & $8.0^{+1.0}_{-1.6}$ & $8.0^{+1.3}_{-1.9}$ & $7.9^{+1.2}_{-1.6}$\\ 
    \covfrac &  $ 0.791^{+0.020}_{-0.023}$ & $0.792^{+0.021}_{-0.020}$ & $0.786^{+0.024}_{-0.020}$ & $0.790^{+0.024}_{-0.018}$\\
    \hline
    \multicolumn{5}{l}{\xstar}\\
    \NH & $5.9^{+2.1}_{-1.7}$ & $5.2^{+2.6}_{-2.3}$ & $4.7^{+2.5}_{-2.2}$ & $4.8^{+2.9}_{-2.0}$\\ 
    \logxi & $3.45^{+0.05}_{-0.06}$ & $3.45^{+0.06}_{-0.07}$ & $3.44^{+0.06}_{-0.08}$ & $3.44^{+0.06}_{-0.07}$\\ 
    \vout$^{a}$  & $4800^{+930}_{-900}$ & $4830^{+1030}_{-900}$ & $5010^{+930}_{-900}$ & $4830^{+1200}_{-1200}$\\ 
    \hline
    \multicolumn{5}{l}{\relxill}\\
    \incl  & $50.7^{+2.9}_{-3.4}$ & $50.8^{+4.0}_{-2.5}$ & $50.1^{+3.0}_{-3.1}$ & $50.3^{+3.0}_{-2.3}$\\ 
    $a$ &  $>0.97$ & $>0.94$ & $>0.93$ & $>0.93$\\ 
    $h$ & $<2.0$ & $<3.2$ & $<2.4$ & $<3.4$\\ 
    $\ringradius$  & $0.0$ & $0.0$ & $<2.1$ & $< 1.8$\\
    $\Gamma$ & $1.75\pm0.08$ & $1.68^{+0.09}_{-0.07}$ & $1.63^{+0.09}_{-0.07}$ & $1.65^{+0.10}_{-0.06}$\\
    \logxi & $3.10^{+0.27}_{-0.17}$ & $2.5^{+0.4}_{-0.5}$ & $2.93^{+0.16}_{-0.27}$ & $2.81^{+0.14}_{-0.47}$\\ 
    $\log n$ & $<17.7$ & $<18.3$ & $<19.1$ & $<18.2$\\ 
    \feabund & $4.0^{+1.6}_{-1.1}$ & $4.7^{+2.5}_{-1.3}$ & $5.2^{+3.4}_{-1.7}$ & $5.1^{+2.8}_{-1.7}$\\ 
    \kTe & $56^{+46}_{-13}$ & $43^{+19}_{-16}$ & $50^{+29}_{-14}$ & $38^{+17}_{-9}$\\
    \boost  & $0.71^{+0.73}_{-0.26}$ & $> 0.26$ & $ > 0.76$ & $1^\mathrm{b}$\\
    \norm & $9^{+10}_{-5}$ & $4.9^{+5.9}_{-4.8}$ & $0.36^{+11.78}_{-0.18}$ & $2.6^{+24.8}_{-2.3}$\\
    \hline
    \multicolumn{5}{l}{\xillvercp}\\
    \norm & $9.6^{+3.3}_{-2.5}$ & $ 8.1^{+3.3}_{-2.3}$ & $7.3^{+3.9}_{-2.2}$ & $6.9^{+3.9}_{-1.8}$\\
    \hline
    \multicolumn{5}{l}{\apec}\\
    \kT & $0.88^{+0.09}_{-0.08}$ & $0.90^{+0.11}_{-0.09}$ & $0.93^{+0.12}_{-0.10}$ & $0.92^{+0.11}_{-0.10}$\\ 
    \norm & $7.4^{+2.2}_{-2.0}$ & $ 6.5^{+2.1}_{-2.0}$ & $5.9^{+2.3}_{-2.0}$ & $6.1^{+2.4}_{-2.0}$\\ 
    \hline
    \multicolumn{5}{l}{\nthcompscat}\\ 
    \norm & $28^{+4}_{-6}$ & $32^{+4}_{-5}$ & $34^{+3}_{-4}$ & $33^{+4}_{-5}$\\ 
    \hline
    $\chi^2$/DoF & 458.0/401 & 458.8/401 & 458.6/400 & 458.8/401\\ 
    \hline
    \hline
    \end{tabular}
\endgroup
\tablefoot{\modellpold corresponds to the lamp post model with constant disk density and no returning radiation, \modellp includes returning radiation and an $\alpha$-disk density gradient. \modelextfreeboost differs from \modellp by varying the new parameter \ringradius, while in \modelextfixboost the \boost parameter is fixed. All uncertainties are given at 90\% confidence. 
Units of $N_\mathrm{H}$ are in $10^{22}\,\mathrm{cm}^{-2}$, the ionization $\xi$ in $\mathrm{erg}\,\mathrm{cm}\,\mathrm{s}^{-1}$, density $n$ in $\mathrm{cm}^{-3}$, inclination $i$ in degrees, height $h$ and radius $\ringradius$ in $\rg$, and model temperatures $kT$ in keV. The normalizations are given in units of $10^{-6}\,\cmminustwo\,\mathrm{s}^{-1}\,\mathrm{keV}^{-1}$, $\mathrm{Norm}_{\relxill}$ is in units of $10^{-4}\,\cmminustwo\,\mathrm{s}^{-1}\,\mathrm{keV}^{-1}$. The iron abundance, $\feabund$ is relative to the interstellar medium abundance of \citet{wilms2000}. a: Outflow velocities are calculated from the redshift parameter $z$ of \xstar as $\vout = (z_\mathrm{cosmological} - z) c$; b: for \modelextfixboost, the \boost parameter is fixed.}
\end{table}

We reproduce \Waltontwentyone's results by applying the lamp post model to the data (\modellpold). Identical to \Waltontwentyone-Model~2a, we use \relxilllpcp\footnote{Note: starting from \relxill~v2.0, \relxilllpcp combines the \textsc{relxill\_ion\_cp} and \textsc{relxill\_d\_cp} flavors that were used in \Waltontwentyone; therefore, the name is different in the aforementioned publication, while the underlying model is the same.} to model relativistic reflection, allowing for a free, but constant disk ionization and density, with no returning radiation. Due to our choice of the optional extraction argument \textit{applyabsfluxcorr=yes,} we found flux cross-calibration constants at ${\sim} 1.25$ and ${\sim} 1.29$ for the \nustar focal plane modules, FPMA and FPMB, with respect to EPIC-pn (Sect.~\ref{subsec:Data}).  Table~\ref{table:parameters_uncertainties_ext} gives all the best-fit parameters.

We found a good fit with $\chi^2_\mathrm{red} = 458.0/401 = 1.142$ and we were able to reproduce the parameters in \Waltontwentyone. The only exception was that we obtained a higher hydrogen column density of the fully and partially covering absorbers, which was expected because we used the abundances \texttt{wilm} instead of \texttt{grsa}. 

Next, we relaxed the assumption on the lamp post by adding returning radiation and self-consistent ionization for an $\alpha$-disk density profile. The latter is especially important for a rapidly spinning BH and a compact primary source \citep[][]{dauser2022}, which is the case for \esosource. This model, \modellp, yields again a good description of the data ($\chi^2_\mathrm{red} = 458.8/401 = 1.144$, see Table~\ref{table:parameters_uncertainties_ext}). Including returning radiation, we find a larger upper limit on the height of the primary source, which increases from $2.0 \rg$ to $3.2 \rg$. The reason is that in the fit the increase in the source height, which decreases the energy shift, is compensated for by the contribution of returning radiation. The ionization gradient, as opposed to constant ionization, results in lower ionization of the disk overall. Instead of an average ionization of $\logxi = 3.10^{+0.27}_{-0.17}$, a maximum ionization\footnote{The maximum ionization of the $\alpha$-disk is at $\diskradius = (11/9)^2 \rin$.} is now $\logxi = 2.5^{+0.4}_{-0.5}$. The limit on disk density, \logdensity, increases inversely proportionally from $\logdensity < 17.7$ to $\logdensity < 18.3$.

Overall, the updated model \modellp does not differ significantly from \modellpold. The ionization, \logxi, is the only parameter that is outside the confidence limit of the previous \modellpold. Therefore, \modellp will serve as a reference for comparison with the extended \relxill model from this work.

\subsection{Ring model}
\label{subsec:results_ext_fits}

\begin{figure}
    \centering
    \includegraphics[width=\columnwidth]{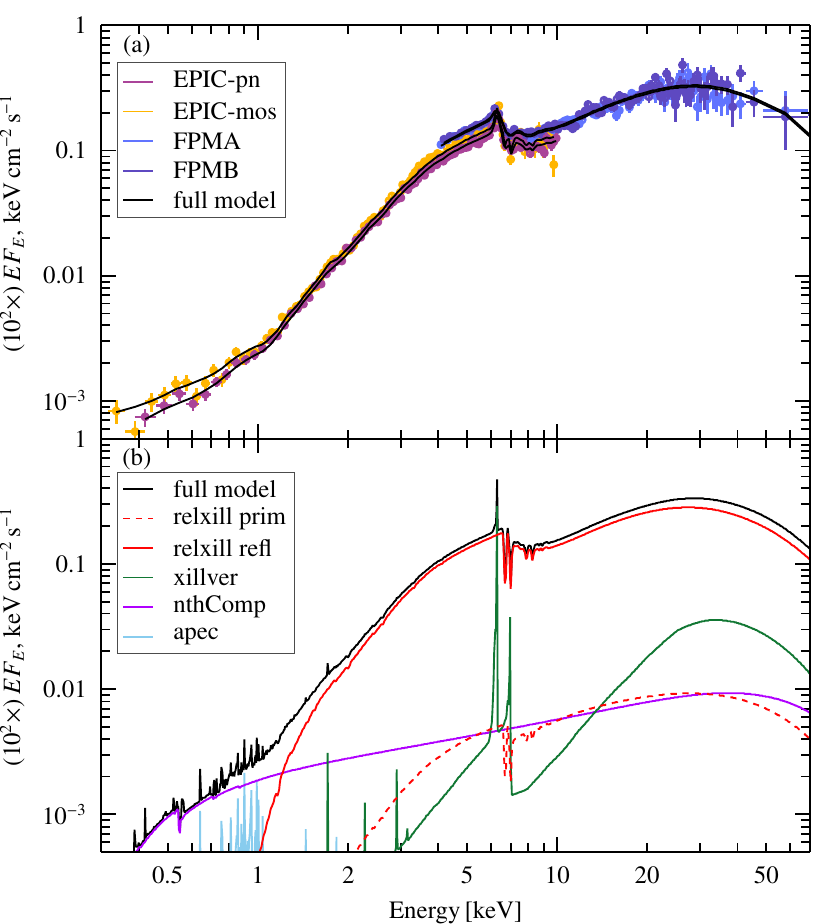}
    \caption{(a) Time-averaged \xmm and \nustar spectrum of \esosource extracted and binned according to Sect.~\ref{subsec:Data} and the full \modelextfreeboost in black. (b)  Components of the best-fit of \modelextfreeboost for \esosource. The parameters of the fit are listed in Table~\ref{table:parameters_uncertainties_ext} and the residuals are shown in Fig.~\ref{fig:resids_ESO033}c.}\label{fig:model_components_ESO033}
\end{figure}

\begin{figure}
    \centering
    \includegraphics[width=\columnwidth]{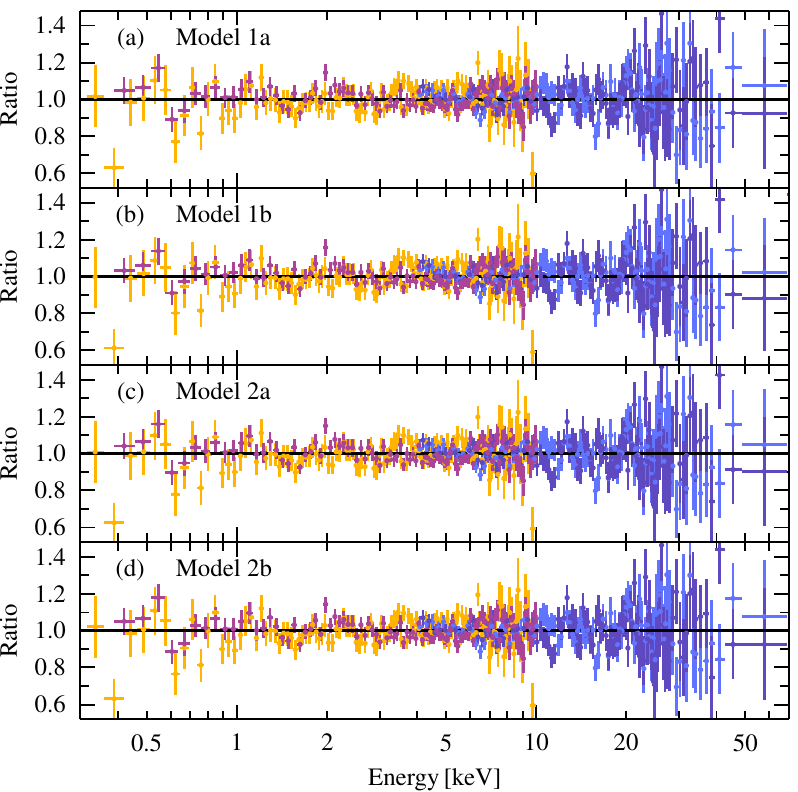}
    \caption{Residuals of the models presented in Table~\ref{table:parameters_uncertainties_ext}. The colors of the data bins have the same meanings as in Fig.~\ref{fig:model_components_ESO033}.}\label{fig:resids_ESO033}
\end{figure}

After finding the lamp post best fit (\modellp), we switched to \relxilllpext and allowed the radial position of the primary source, \ringradius, to vary freely (\modelextfreeboost). The best fit model and residuals are shown in Fig.~\ref{fig:model_components_ESO033} and Fig.~\ref{fig:resids_ESO033} (see Table~\ref{table:parameters_uncertainties_ext} for all fit parameters). Relativistic reflection is the main contribution to the model at energies ${\gtrsim}1$\,keV, except for the iron line at 6.4\,keV, which originates in the distant reflector. The primary flux component of \relxill is about an order of magnitude lower than the reflected component, due to the compact primary source geometry and rapidly rotating BH, $a > 0.93$. 

We found that the new ring-like model describes the data equally well ($\chi^2/\textrm{DoF} = 458.6/400$) and we found no major differences to the lamp post model in the residuals. The radius of the ring-like primary source is constrained below $\ringradius < 2.1\rg$, excluding a primary source with a larger radius. This result is expected since small values of \ringradius create line profiles similar to the lamp post (see Fig.~\ref{fig:relline_spectra}). The height is constrained to ${<}2.4 \rg$. All other parameters are comparable to \modellp within uncertainties. 

Additionally, we test \modelextfixboost with the normalization of the reflected and primary spectrum fixed to the one expected from a ring with an isotropic emission pattern (i.e., we fix the parameter $\boost = 1$). The results are also shown in Table~\ref{table:parameters_uncertainties_ext}. We do not find significant differences between the two models. The reason being that the spectrum is very reflection dominated, which can also be seen in the fact that \modelextfreeboost only sets a lower limit on the boost parameter of $\boost > 0.76$. We note that while \boost values ${\gg} 1$ are formally allowed by the fit, they can be ruled out as non-physical because such values would require the primary emission to be exceedingly boosted towards the disk. In Fig.~\ref{fig:resids_ESO033} we show the residuals of both models, \modelextfreeboost and \modelextfixboost, compared to the fits of the lamp post model. There are no significant differences in the residuals between the models. 

\begin{figure}
    \centering
    \includegraphics[width=\columnwidth]{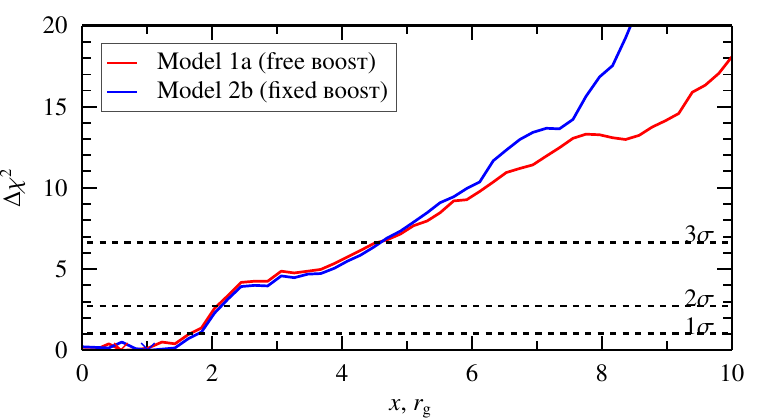}
    \caption{ 1D~confidence contours for the radius of the primary source, \ringradius, calculated for \modelextfreeboost and for \modelextfixboost. The horizontal dashed lines show the $1\sigma$, $2\sigma$, and $3\sigma$ confidence levels for \ringradius values.}\label{fig:steppar_ring}
\end{figure}

We show the confidence contours for the radius of the primary source in Fig.~\ref{fig:steppar_ring}, comparing \modelextfreeboost and \modelextfixboost. Despite the stronger assumption of $\boost = 1$ in \modelextfixboost, both models show a similar behavior of a slowly increasing $\Delta \chi^2$ with the ring radius. Only for large source radii, a free normalization between the reflection and the direct spectrum leads to a weaker constraint on the radius. These results confirm that there is no second solution with a large source radius that would describe the data equally well or better than a compact lamp post-like source. 

\section{Discussion and conclusions}
\label{sec:general_results}

We have developed a new reflection model that describes X-ray reflection for a ring-like source on the symmetry axis of a rotating BH. The emission line profiles and properties of the reflection spectrum indicate (not unsurprisingly) that in the limit of small ring radii, only minor differences are found between the ring-like source and a point source. At large source radii, significant differences in the irradiation profile and in the line profile are present (Sect.~\ref{sec:results_model}). This means that it is, in principle, possible to constrain the size of the irradiating medium through X-ray spectral modeling. 

In this work, we applied our model to a well-studied observation of \esosource (Sect.~\ref{sec:results_data}). In the previous study \Waltontwentyone showed that the reflection spectrum can be described by a lamp post. Here, after applying the extended version of \relxill,  an upper limit was found for the radius of the ring-like source of primary X-rays. Specifically, we found $h < 2.4 \rg$ and $\ringradius < 2.1 \rg$, We can also exclude the existence of a more extended ring-like source. 
To our knowledge, this is the first (or at least among the first) direct constraints set on the size of the source of hard X-rays that is solely based on X-ray spectroscopy \citep[noting that other size constraints can be obtained, e.g., from X-ray timing analysis, see][]{cackett2021}. The availability of the new version of \relxill means that a significant limitation of earlier reflection models, namely, the assumption of small coronal sizes, has been lifted. This implicit assumption of very small primary source sizes has been one of the major criticisms of lamp post modeling in the literature.

In this section, we address several important aspects and caveats that need to be kept in mind when applying the ring model. We start in Sect.~\ref{sec:interpretation_of_lp} with an in-depth discussion of the consequences of our initial result of a small corona for accretion physics, followed by discussions of potential parameter correlations (Sect.~\ref{subsec:results_parameters_mcmc}) and our assumption that the primary source has rotational symmetry (Sect.~\ref{sec:timescales}). We summarize our results in Sect.~\ref{sec:summary}.

\subsection{Lamp posts and ring-like geometries}
\label{sec:interpretation_of_lp}

The lamp post geometry is a very rudimentary description of a compact spherical primary source on the rotational axis of the BH, described as a point source with a spherically isotropic emission, which is only defined by its height above the BH. Despite these limitations, relativistic reflection models using the lamp post geometry have been very successful in describing observational data of Galactic BH binaries and AGNs. It has been generally assumed that this means that the astrophysical corona is also small. However, before the availability of the model discussed here, there were no tools available that would allow for reflection spectroscopy to be used as a test of the presence of an extended primary source geometry. While a detailed assessment of the validity of the lamp post model is beyond the scope of this study, we summarize the overall behavior below. We note that this discussion should only put earlier fits with the lamp post geometry in perspective. For more reliable estimates of the source geometry, earlier data sets will have to be re-analyzed with a reflection model including a radially extended source.

From the line profiles (see Fig.~\ref{fig:relline_spectra} and Sect.~\ref{sec:results_model_iron}), we can see that the distance to the BH, \sphericalradius, is mainly influencing the relativistic line broadening; therefore, the shape of the relativistic reflection spectrum for a compact primary source. The position in terms of polar angle only has a minor effect on the line shape. A change of this angle does not produce significantly different line shapes, but can likely be compensated by slightly different values of \sphericalradius. This means that reflection from a point-like source located on the rotational axis produces a similar relativistic line broadening to a radially extended source at a lower height above the disk, but at a similar distance to the BH. Even taking into account the full reflection spectrum including the primary emission, Fig.~\ref{fig:relxill_refl_spectra} shows that for a standard lamp post height of $h=3\rg$, there are only weak differences between the lamp post and a ring source with $\ringradius < 0.5 \risco$ modulo normalization.
 
These general conclusions have been confirmed by our analysis of the data of \esosource (Sect.~\ref{sec:results_data}), where we find that a ring-like geometry with radius $\ringradius < 2.1\rg$ gives consistent results to the lamp post geometry. We therefore expect that observations that have been described with lamp post models of a small height are also compatible with ring sources, provided that the ring is only extended by a few gravitational radii. Other extragalactic sources falling in this range include Mrk\,335 \citep{parker2014}, 1H0707$-$495 \citep{kara2015}, NGC\,4151 \citep{beuchert2017}, and IRAS\,09149$-$6206 \citep{walton2020}.

A minimum estimate for the size of the primary source is also necessary for the primary source to be energetically stable and to ensure that enough seed photons from the accretion disk intercept the primary source to allow for a Compton upscattering of the observed spectrum \citep{dovciak2016}. By applying this constraint to \esosource, \Waltontwentyone found that the radius of the spherical source must be at least ${\sim}0.2\rg$. Although our own model assumes a ring instead of a sphere, with our upper limit of the radius of $\ringradius < 2.1\rg$, our results also agree well with this limit as long as the ring is slightly extended.

\subsection{Parameter correlations}
\label{subsec:results_parameters_mcmc}

\begin{figure*}
\sidecaption
    \includegraphics[width=12cm]{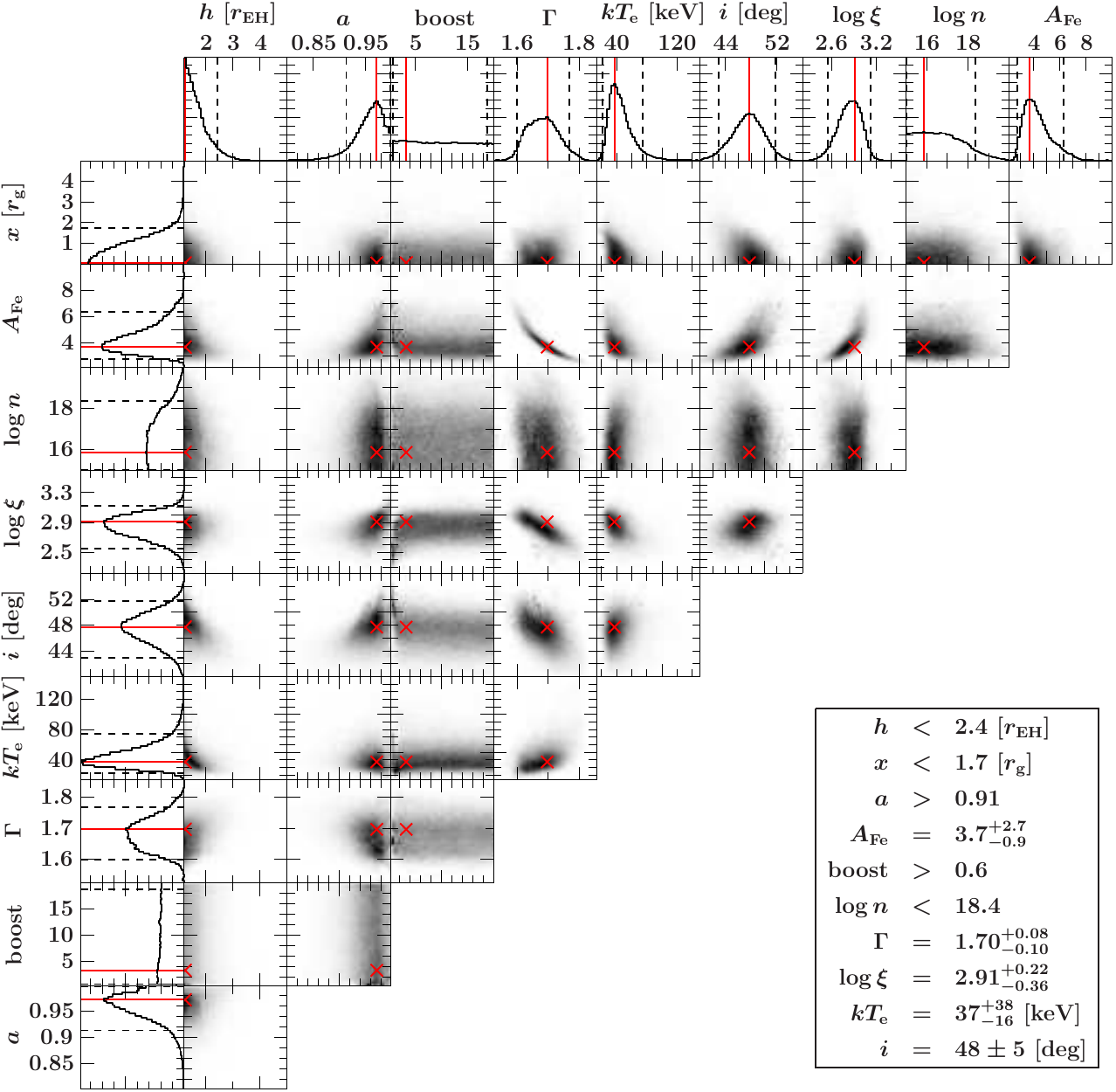}
    \caption{MCMC simulation results for the relevant parameters of \modelextfreeboost. Each 1D histogram shows probability distribution of a single parameter values. Each 2D histogram shows probability distribution of each pair of parameters. The red crosses mark the most probable parameter positions according to 1D probability distributions. Additionally, the inset in the lower right corner shows the most probable values with 90\% confidence intervals.}\label{fig:MCMC_ESO033}
\end{figure*}

The analysis of real source data requires us to understand potential correlations between parameters of the model, since they could influence the interpretation of the results. We did so by again using the example of the \xmm and \nustar observation of \esosource discussed in Sect.~\ref{sec:results_data}. Specifically, we study the parameter space through probability distributions of the parameters obtained from Markov chain Monte Carlo (MCMC) sampling \citep{foreman-mackey2013}. We used ten walkers for 40000 steps and remove 5000 first steps as a ``burn-in'' phase. In total, we used 350000 parameter combinations. Figure~\ref{fig:MCMC_ESO033} shows the result of MCMC runs\footnote{For all MCMC runs, we set the environment variable $\texttt{RELXILL\_RENORMALIZE}=1$, available since \relxill v2.4, which redefines the normalization of the total flux to have the same value at 3\,keV at each model evaluation. This approach removes the degeneracy of normalization with other parameters (such as the height) and helps the MCMC to converge faster. } for \modelextfreeboost. Generally, the MCMC results agree well with the best fit values presented in Table~\ref{table:parameters_uncertainties_ext}. 

The MCMC runs reveal a strong anti-correlation of the inferred iron abundance, \feabund, and the photon index, $\Gamma$, and degree of ionization, \logxi. These correlations are not unexpected, and similar to the well-known correlation between $\Gamma$ and the foreground absorption column, $\NH$. Specifically, since the reflection spectrum increases towards higher energies, a softer spectral shape (i.e., with increasing $\Gamma$) can be compensated for by the contribution of the reflection continuum at lower energies. This can be done by increasing $\feabund$, since this changes the contribution of the reflection spectrum at lower energies, especially in combination with the change in $\log\xi$. In addition, a higher disk density (and therefore a lower ionization) also increases the reflected flux at low energies, such that the overall spectral shape appears softer. A similar anti-correlation was found by \citet{tomsick2018}.

\begin{figure}
\centering
    \includegraphics[width=\columnwidth]{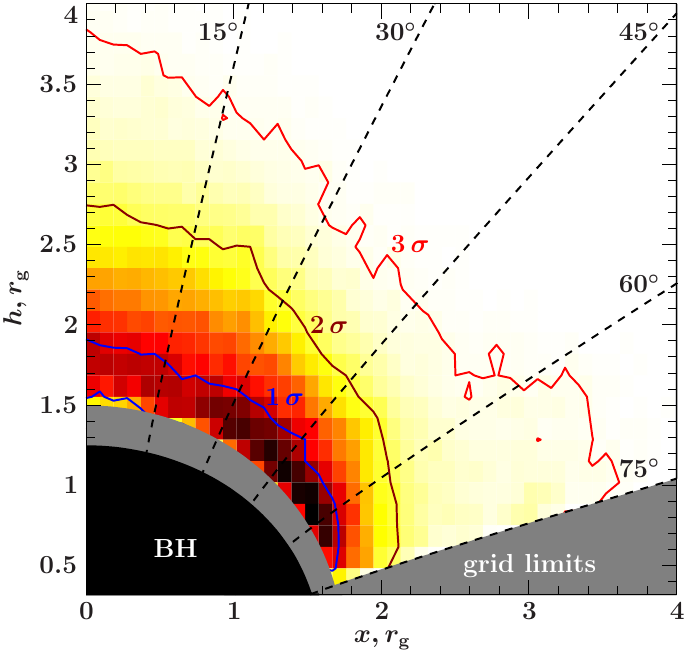}
    \caption{Ring size and height probability distribution with $1\sigma$, $2\sigma$, and $3\sigma$ confidence levels, based on the MCMC results (Fig.~\ref{fig:MCMC_ESO033}, for \modelextfreeboost) extended to lower heights. The color scale shows the probability distribution. The dashed lines show combinations of $h$ and $\ringradius$ corresponding to constant polar angles, $\theta$. %Dotted lines show the same for constant $\sphericalradius$. 
    The black region is limited by the BH event horizon, the gray region denotes \relxill parameter grid limits. 
    }\label{fig:MCMC_contour_levels}
\end{figure}

The MCMC runs allowed us to understand the correlation between the source's geometrical parameters. In Fig.~\ref{fig:MCMC_contour_levels}, we show a more detailed version of the MCMC probability distribution, including lines showing polar coordinates of constant distance to the BH, \sphericalradius, and angular position, $\theta$, of the source. The confidence contours roughly align with the distance of the ring to the BH, \sphericalradius, which has to be ${<}2.7\rg$ at high significance regardless of the height or radius of the ring, $(h, \ringradius)$. In contrast, the position of the source from $0^\circ$ to $75^\circ$ is relatively unconstrained. While we can constrain the distance and, therefore, the compactness of the primary source in \esosource very well, its actual location is still not well constrained. The source can be both a ``lamp post-like'' source with a very small radial extent, as well as a ``ring-like'' source with larger radius at a lower height. This result emphasizes the importance of the distance to the BH in general, regardless of the polar angle.

Finally, we note that the probability distribution of the \boost parameter is almost flat and we were only able to obtain a lower limit of $\boost \gtrsim 0.6$. This result indicates that the primary component is not required in the fit and, therefore, the data can be solely described by reflection. This is expected because in the extreme configuration of a very compact primary source, most of the radiation is focused toward the disk and reflected \citep{dauser2014}. With returning radiation included in the model, this effect becomes even more drastic. Similarly, we did not find any significant correlations between the ring radius, $\ringradius$, and other parameters. Importantly, $\ringradius$ is also not correlated with the BH spin, $a$, such that any combination of small values of $h$ and $\ringradius$ is compatible with the observed data.

\subsection{Model limitations}
\label{sec:timescales}

Throughout this work, we assumed an axisymmetric primary source. However, intrinsically, the source may be asymmetric or even be an off-axis source rotating around the BH. The resulting spectral variations will be averaged over the exposure time if we apply the reflection model to observations that are longer than the Kepler timescale at the ISCO, $T_\mathrm{Kepler}$. To first order, for a maximally rotating BH, we have
\begin{equation}
T_\mathrm{Kepler} = 
\frac{2 \pi \mathrm{G} M}{\mathrm{c}^3}=31\,\mathrm{s} \left(\frac{M}{10^6\,\mathrm{M}_\odot}\right)
,\end{equation}
so that for most realistic, ks-long, exposures the averaging condition is fulfilled.

We also emphasize that the ring geometry used here is still an approximation of a more complicated spatial distribution of matter that we would expect in more realistic models for the inner accretion flow close to the BH. It is important to stress that (similarly to the results of lamp post models), the specific values for the ring radius or height should not be overinterpreted and should only be taken as approximations of the spatial dimensions of even more realistic shapes of the accretion flow. 

\subsection{Conclusions}\label{sec:summary}

We present a new relativistic reflection model with a ring-like primary source of X-rays. The model has been integrated into the publicly available \relxill reflection model. We find that the main factor influencing the shape of the reflection spectra is the distance of the primary source to the BH, while the polar angle of the source has a weaker effect (Sect.~\ref{sec:results_model_iron}). This conclusion applies especially for compact sources with ring radii of a few $\rg$ where the spectral shape stays similar to the lamp post one, but the reflection fraction is increased (Sect.~\ref{sec:results_model_continuum}). This means that if the true corona is extended, while lamp post models will still describe the data, they will end up overestimating the reflection fraction (Sect.~\ref{sec:interpretation_of_lp}).

By constraining the radius of the primary source for \esosource for the first time, we find a compact geometry of the primary source with a vertical location below ${\sim} 3 \rg$ and radial location below ${\sim}2 \rg$ at a 90\% confidence level (Sect.~\ref{subsec:results_ext_fits}). We also show that any source at a distance of ${<}2.7 \rg$ to the BH provides an equally good description of the data, with only a weak dependence on the polar angle in a wide range $[0^\circ, 75^\circ]$. 

The more physical representation given by the ring geometry provides us with the opportunity to constrain the size of the primary source. Although the ring source geometry is still overly simplified, it can serve as a basis for further development of more general axisymmetric extended primary source models. For example, slab-like, conical, or spherical primary sources can be easily constructed by combining several emitting rings of different radii and heights. To make further progress, knowledge about the actual geometry of matter is crucial for our understanding of the processes taking place in that compact space-time region. One possible interpretation is that the primary source is connected to the jet launching region, which would provide a connection between the jet base and the accretion flow \citep[e.g.,][]{markoff2005}. We still caution that the real structure of the jet base is expected to be much more complicated than that of a single ring source \citep[e.g., see][for a review of GRMHD simulations with thin disks]{dihingia2024a}. However, the ring model already allows us to fit the radius of the primary source in real time. 

Since relativistic reflection spectra mainly depend on the distance to the BH (Sect.~\ref{sec:interpretation_of_lp}), any spectral fits based on a lamp post geometry are likely compatible also with a source that is radially extended by a few $\rg$. Recent results using X-ray polarimetric data taken by the Imaging X-ray Polarimetry Explorer \cite[IXPE,][]{weisskopf2022} of a few selected bright Seyfert galaxies, such as NGC\,4151 \citep{gianolli2023} and IC\,4329A \citep{ingram2023}, have shown that the medium responsible for creating the polarization has to be roughly perpendicular to the radio jet. These results have been interpreted as evidence of a likely radially extended primary source, as opposed to the compact spherical lamp post that has been employed over the last decade to explain most relativistic reflection spectra. 
However, these apparent differences between the inferred primary geometries from  reflection and polarization studies can be reconciled in light of the results presented here. Furthermore, this would also support the X-ray timing results, which generally prefer small time travel distances between the source and the disk \citep[e.g.,][]{kara2016,cackett2021}. The best constraints on the primary source geometry will therefore require a model that includes X-ray polarimetry and reflection for a radially extended primary source. 

\begin{acknowledgements}
The authors thank the anonymous referee for the detailed and constructive report with highly useful suggestions on improving the manuscript content. This research is supported by the DFG research unit FOR 5195 `Relativistic Jets in Active Galaxies' (project number 443220636, grant number WI 1860/20-1). DJW acknowledges support from the Science and Technology Facilities Council (STFC; grant code ST/Y001060/1). The results in this paper are in part based on observations obtained with XMM-Newton, an ESA science mission with instruments and contributions directly funded by ESA Member States and NASA. This research has made use of data from the NuSTAR mission, a project led by the California Institute of Technology, managed by the Jet Propulsion Laboratory, and funded by the National Aeronautics and Space Administration. Data analysis was performed using the NuSTAR Data Analysis Software (NuSTARDAS), jointly developed by the ASI Science Data Center (SSDC, Italy) and the California Institute of Technology (USA).  This research has made use of ISIS functions (ISISscripts) provided by ECAP/Remeis observatory and MIT (\href{http://www.sternwarte.uni-erlangen.de/isis/}{www.sternwarte.uni-erlangen.de/isis/}).
\end{acknowledgements}

\bibliographystyle{aa}
\bibliography{bibliography}

\begin{appendix}

\section{Energy shift formula}
\label{app:energyshift}

\subsection{Energy shift from source to disk}
    
Following \citet{niedzwiecki2005} and \citet{bardeen1972}, we write the photon energy in the locally non-rotating rest frame (LNRF) as
\begin{equation}\label{eq:energy_lnr_frame_appendix}
    \Elnrf = \Einf \left( 1 - \omega \lambda\right) e^{-\nu}\;,
\end{equation}
where $\omega$, $e^{-\nu}$ are defined in Eq.~\eqref{eq:coefficients_metric}, $\Einf$ is the photon energy measured at infinity and $\lambda$ is a constant of motion, defined through the angular momentum component of the photon parallel to the rotational axis of the BH. 
    
The photon energy in the rest frame of the circularly rotating plasma (i.e., a point of the disk or a point of the ring source in this calculation) following a Lorentz transformation is expressed as \citep{niedzwiecki2005}:
\begin{equation}\label{eq:energy_rest_frame_appendix}
    \Erest = \gamma \left(\Elnrf - v^{\varphi} p_{\varphi}\right),
\end{equation}
where $v^{\varphi}$ is the physical azimuth velocity of the plasma, defined by Eq.~\eqref{eq:velocity}, $\gamma = \left(1 - {v^{\varphi}}^2\right)^{-1/2}$ is the corresponding Lorentz-factor, $p_{\varphi}$ is the azimuthal component of 4-momentum of the photon, which can be written as \citep{niedzwiecki2005}
\begin{equation}\label{eq:pphi_appendix}
    p_{\varphi} = \frac{\Einf \lambda}{\sin\theta} \sqrt{\frac{\Sigma}{A}}\;,
\end{equation}
where $A$, $\Sigma$ are defined in Eq.~\eqref{eq:coefficients_metric} and $\theta$ is the polar angle from the metric, Eq.~\eqref{eq:kerr_metric}. Using Eqs.~\eqref{eq:velocity} and~\eqref{eq:pphi_appendix}, the term $v^{\varphi} p_{\varphi}$ in Eq.~\eqref{eq:energy_rest_frame_appendix} can be rewritten in explicit form
\begin{equation}
    v^{\varphi} p_{\varphi} = \Einf e^{-\nu} \left(\OmegaSource - \omega\right) \lambda\;,
\end{equation}
such that
\begin{equation}\label{eq:energy_shift_disk_inf_appendix}
    \frac{\Einf}{\Erest} = \frac{1}{\gamma e^{-\nu} \left(1 - \OmegaSource \lambda\right)}\;,
\end{equation}
which defines the energy shift from the disk to the infinity \citep[also in][]{niedzwiecki2005}. We can already use Eq.~\eqref{eq:energy_shift_disk_inf_appendix} to define the energy shift of the photon propagating from the ring source to infinity, if we take all the quantities $e^{-\nu}$, $\OmegaSource$, $\gamma$ at the location of the source.
    
For a given photon with energy $\Einf$ at infinity, Eq.~\eqref{eq:energy_shift_disk_inf_appendix} allows us to obtain the equation for the energy shift of the photon between the rest frames of the primary source and the disk, 
\begin{equation}\label{eq:energy_shift_appendix}
    g = \frac{\left.\Erest\strut\right|_\mathrm{disk}}{\left.\Erest\strut\right|_\mathrm{source}} = \frac{\left.\Erest\strut\right|_\mathrm{disk}}{\Einf} \frac{\Einf}{\left.\Erest\strut\right|_\mathrm{source}} = \frac{\left[\gamma e^{-\nu} \left(1 - \OmegaKeplerian \lambda\right)\strut\right]_\mathrm{disk}}{\left[\gamma e^{-\nu} \left(1 - \OmegaSource \lambda\right)\strut\right]_\mathrm{source}}\;,
\end{equation}
where all the quantities $e^{-\nu}$, $\OmegaKeplerian$, $\OmegaSource$, $\gamma$ are taken at the location in the disk, or at the source location, when the corresponding subscript is given. For $\lambda = 0$ and a source located at the rotational axis of the BH, Eq.~\eqref{eq:energy_shift_appendix} reduces to the lamp post energy shift \citep{dauser2013}.

\subsection{Lensing and energy shift from source to observer}
\label{subsec_app:lensing}

The flux propagating directly from the source to the observer is affected by the energy shift and by the lensing, namely, the effect of redistribution of the flux propagating in non-Euclidean metric. The energy shift experienced by the photons propagating from the source directly to the observer is already given by Eq.~\eqref{eq:energy_shift_disk_inf_appendix}, $\lambda$ is the only value of that expression which requires raytracing. All other values are known, and we follow the same procedure of energy shift binning as explained in Sect.~\ref{subsec:implementation} for the case of the energy shift source-disk.

Lensing defines how the photon flux distribution on the sky changes as photons propagate from the source to the observer. For lensing of the flux, we adopt the lensing calculation procedure for the lamp post source described by \citet{ingram2019a} for our case of the extended source \citep[see also][]{feng2025}. In that publication, lensing was defined through the angle derivatives in terms of inclination, or the final polar angle of the photon (the observer sky) and the angle of the emitted photon (the source sky). We take a step back and write the definition of lensing in terms of a ratio of the fluxes in a given patch of the sky, or, equally, solid angles in the sky as 
\begin{equation}\label{eq:lensing}
    l = \frac{\dif F_\mathrm{o}}{\dif F_\mathrm{s}} = \frac{\dif \Omega_\mathrm{o}}{\dif \Omega_\mathrm{s}} = \frac{\dif \cos\theta_\mathrm{o}\dif \varphi_\mathrm{o}}{\dif \cos\theta_\mathrm{s}\dif \varphi_\mathrm{s}}\;,
\end{equation}
where $\dif F_\mathrm{o,s} = I \dif \Omega_\mathrm{o,s}$ are fluxes from a given patch of the source and the observer sky, $I$ is the constant intensity, $\dif \Omega_\mathrm{s}$, $\dif \Omega_\mathrm{o}$ are solid angle derivatives in the source sky and the observer sky \citep[][their Fig.~1]{ingram2019a}, and $(\varphi_\mathrm{s}, \theta_\mathrm{s})$, $(\varphi_\mathrm{o}, \theta_\mathrm{o})$ are the spherical coordinates in the source sky and the observer sky, respectively, and $\theta_\mathrm{o}$ can be chosen equal to the observer's inclination, $i$. Equation~\eqref{eq:lensing} is valid as long as the photon intensity, $I$, is conserved. 

As the ring source is axially symmetric, only the flux distribution for different inclinations matters for the observer. Therefore, we divide the sky of the observer into sectors $\Delta \Omega_\mathrm{o} \colon \left[0, 2 \pi\right) \times \left[\cos i, \cos i + \Delta \cos i\right)$, where $i = \theta_\mathrm{o}$ is the inclination and $\Delta \cos i$ is the step of the cosine of inclination, which we take constant to keep a comparable number of photons for all inclinations in the range $[0, \pi/2]$. 

After ray-tracing, each sector contains a certain amount of photons. The same photons have different positions in the source sky, and each sector is ``projected'' into the source sky. After that, the lensing calculation reduces to a comparison of the areas taken by sectors in the source sky and the observer sky, $l \approx \Delta \Omega_\mathrm{o} / \Delta \Omega_\mathrm{s}$. Calculating $\Delta \Omega_\mathrm{o} = 2 \pi \Delta\cos i$ is trivial, while the area of the sector in the source sky $\Delta \Omega_\mathrm{s}$ can be calculated by taking the integral of each sector. We additionally use the fact that the sector area relates to the total area taken by photons reaching infinity as the number of photons relates to the total number of escaping photons (aka Monte Carlo integration). Therefore, $\Delta \Omega_\mathrm{s} = 2 \pi \text{[photons in sector]} / \text{[total photons]}$.

\section{Details of the implementation of the algorithm into \relxill}
\label{app:implementation}

We divide the disk into $K$ radial circular bins, that is, annuli described in Sect.~\ref{subsec:disk_irradiation}, with inner edges of each bin at $r_k$, $k = \overline{0,K-1}$, where $r_0$ corresponds to the inner edge of the disk, $r_0 = \rin$, and $r_{K-1} = \rout$ corresponds to the bin reaching the outer edge of the disk. The width of the bin is $\Delta r_k = r_{k+1} - r_k$. We choose the optimal distribution of the radial bins, $r_k$, to keep a sufficient statistical distribution of photons in each bin in the whole range of parameters $(a, h, \ringradius)$. The parameters of the binning are the following: the total number of bins $K = 100$, with $80$ bins distributed $\propto \diskradius^2$ below $100 \rg$ and $20$ bins $\propto \diskradius^{1.5}$ above $100 \rg$.

For the chosen binning and velocity profile, defined by Eq.~\eqref{eq:velocity}, we simulate as many photons as possible to improve the statistics of our simulations. The main limitation to get a smooth statistical distribution of photons is the computational time required to cover all combinations of the geometric parameters (spin, ring geometry), so we find and use the optimal number of photons $N = 10^8$ for each simulation. 

We define the range of the geometric parameters in which we simulate the table values. We simulate fluxes in the whole physically acceptable range of spins $-0.998 \leq a \leq 0.998$ with $20$ spin values spread over this range. As noted in the main text, there are different possibilities to define the ring position, either heights and ring radii $(h, \ringradius)$ or spherical radii and polar angles $(\sphericalradius, \theta)$. We produce tables in both coordinate systems. For heights, we select the range $1.1\reh \leq h \leq 50\rg$, where $\reh$ is the radius of the event horizon of BH, with $50$ values distributed in this range, and with a possible extension of the height range further down to $h = 0.2\rg$. For spherical radii, we also select the range $1.1\reh \leq \sphericalradius \leq 50\rg$. For the ring radii, we select the range $0 \rg \leq \ringradius \leq 50 \rg$ with $50$ values distributed in the given range. For angles, we use $0^\circ \leq \theta \leq 80^\circ$ with step of $5^\circ$. We note again that $\ringradius = 0$ or $\theta = 0^\circ$ corresponds to the lamp post model. We use $(\sphericalradius, \theta)$ to produce Figs.~\ref{fig:relline_spectra} and~\ref{fig:MCMC_contour_levels} and find that this representation performs better with common fitting algorithms and MCMC. In total, we have a grid of parameters $(a, h, \ringradius)$ with $20 \times 50 \times 50$ combinations or $20 \times 50 \times 17$ for $(a, \sphericalradius, \theta)$. In addition to fluxes and energy shifts, we store the photon fractions, \fractiondisk, \fractionesc, \fractionblackhole, to calculate the normalization between the reflected and direct spectrum.
    
For direct flux normalization, we also tabulate lensing and energy shift values (Appendix~\ref{subsec_app:lensing}) for different geometric parameters and inclinations. We take $20$ uniform inclination sectors in the range $[0, \pi/2]$, obtain and tabulate average lensing inside of each inclination sector, and tabulate energy shifts in the same way as the energy shifts of the photons propagating to the disk.

Proper relation of the reflected flux to the direct emission of the primary source is important for the analysis of relativistic reflection. In \relxill this relation is defined through reflection fraction, \reflfrac, parameter \citep[see][for details]{dauser2016}, or, equivalently, \boost parameter, which is defined as the ratio of the measured reflection fraction to the reflection fraction predicted from the geometry of the isotropically emitting primary source. In order to calculate \reflfrac or \boost parameter, we have to consistently normalize the irradiating flux. We use the following scheme to normalize the flux. Far from the BH, $\diskradius \to \infty$, the disk irradiation flux tends to non-relativistic limit in flat space-time, $F_\mathrm{flat}$. We set the normalization of the relativistic flux according to this non-relativistic limit. For that, we use the simple fact that in non-relativistic limit the isotropic photon flux produced by a number of photons $N$, integrated over the infinite disk would always be $\int F_\mathrm{flat} \dif\flatarea = 0.5 N$. Therefore, we normalize the irradiating flux, Eq.~\eqref{eq:irradiating_flux}, dividing it by $0.5 N$. This approach gives the same normalization as the lamp post profile of \relxill at a given height, when $\ringradius = 0$. At the same time, with varying ring radius and fixed height, any radius in the disk, $\diskradius$, far enough from the primary source, has the same photon intensity value, as the lamp post profile at the same height. This is because the ring source becomes indistinguishable from the point source if $\diskradius \gg \ringradius$.

\section{Comparison to other models}
\label{app:comparison}

\subsection{\textsc{relxill\_nk}}

\textsc{relxill\_nk} \citep{abdikamalov2019} is a relativistic reflection model based on \relxill and focusing on studies of non-Kerr metrics. \citet{riaz2022} implemented ring and disk primary source geometries, with disk irradiation tables available upon request (at the time of writing). 

Several differences in implementations make the detailed comparison impractical. Most importantly, we raytrace photons in all spatial directions (Sect.~\ref{subsec:raytracing}), while \textsc{relxill\_nk} does so only for twelve vertical 2D-slices perpendicular to the disk. Moreover, the code does not account for the variation of energy shifts of photons propagating in different directions (see Sect.~\ref{subsec:energy_shift_spread}). Less importantly, the definition of height and ring radius in \textsc{relxill\_nk} does not account for spin dependence (compare Eq.~\ref{eq:metric_to_hx} and Eq.~2 in \citealp{riaz2022}). With given differences, \textsc{relxill\_nk} with ring geometry predicts up to a few ten percent more redshift to the broad line profiles, compared to our model. 

\subsection{\textsc{reflkerr}}

\textsc{reflkerr} model by \citet{niedzwiecki2019} has only cylindrical geometry publicly available for fitting with relativistic reflection model \citep{szanecki2020}, and it predicts only the full reflection spectra (i.e., no line model publicly available). In addition, \textsc{reflkerr} uses different primary continuum prescription, making a comparison more complicated. Therefore, a direct comparison with ring geometry is not possible. Moreover, while comparing the \textsc{reflkerr} output for different parameters, we do not find a smooth transition in the predicted flux when the size of the primary source reaches zero in height and radius (i.e., recovering the lamp post geometry, available separately in \textsc{reflkerr} package). In this case, we find a sharp increase in the simulated flux.

\subsection{\textsc{Gradus.jl}}

We find a good agreement with the disk irradiation profiles and line profiles produced with \textsc{Gradus.jl} \citep{baker2025}\footnote{\href{https://github.com/astro-group-bristol/Gradus.jl}{https://github.com/astro-group-bristol/Gradus.jl} or \href{https://ascl.net/2503.035}{https://ascl.net/2503.035}}. The difference in our profiles is a few percent or less, caused by the numerical noise of two different implementations. The transfer functions of \textsc{Gradus.jl} and \relxill agree at a level better than one percent. Subsequently, the line profiles have differences of the same order of magnitude as the disk irradiation profiles, decreasing for larger heights of the corona. We actively collaborate on further improvements of the precision of the two very distinct codes attempting to explain the same phenomena.

\end{appendix}

\end{document}